\documentclass[]{spie}  

\usepackage{amsmath,amsfonts,amssymb}
\usepackage{graphicx}
\usepackage[colorlinks=true, allcolors=black]{hyperref}
\usepackage{subfig}
\usepackage{wasysym}

\title{Recent developments in X-ray lens modelling with SRW}

\author[a]{Rafael Celestre}
\author[b]{Oleg Chubar}
\author[a]{Thomas Roth}
\author[a]{Manuel Sanchez del Rio}
\author[a]{Raymond Barrett}
\affil[a]{ESRF - The European Synchrotron, 71 Avenue des Martyrs, 38000 Grenoble, France}
\affil[b]{National Synchrotron Light Source II, Brookhaven National Laboratory, NY 11973, USA}

\authorinfo{Further author information: rafael.celestre@esrf.eu\\ }

\begin{document}

\maketitle
\begin{abstract}
The advent of 4$^\text{th}$ generation high-energy synchrotron facilities (ESRF-EBS and the planned APS-U, PETRA-IV and SPring-8 II) and free-electron lasers (Eu-XFEL and LCLS-II) allied with the recent demonstration of high-quality free-form refractive optics for beam shaping and optical correction have revived interest in compound refractive lenses (CRLs) as optics for beam transport, probe formation in X-ray micro- and nano-analysis as well as for imaging applications. Ideal CRLs have long been made available in the 'Synchrotron Radiation Workshop' (SRW), however, the current context requires more sophisticated modelling of X-ray lenses. In this work, we revisit the already implemented wave-optics model for an ideal X-ray lens in the projection approximation and propose modifications to it as to allow more degrees of freedom to both the front and back surfaces independently, which enables to reproduce misalignments and manufacturing errors commonly found in X-ray lenses. For the cases where simply tilting and transversely offsetting the parabolic sections of a CRL is not enough, we present the possibility of  generating the figure errors by using Zernike and Legendre polynomials or directly adding metrology data to the lenses. We present the effects of each new degree of freedom by calculating their impact on point spread function and the beam caustics. 
\end{abstract}
\keywords{X-ray optics, CRL, X-ray lens, optical modelling, physical optics, SRW}

\section{INTRODUCTION}\label{sec:intro}

Refractive X-ray optics comprises mainly lenses [\cite{Tomie1994, Snigirev1996}], prisms in several arrangements [\cite{Cederstrom2000, Jark2004}] and most recently, free-form objects mainly for optical correction and beam-shaping [\cite{Seiboth2017, Zverev2017, Seiboth2019, Seiboth2020}]. From those, parabolic compound refractive lenses are by far the dominating refractive optical element in use throughout synchrotrons, hence in this work emphasis is given in describing typical misalignment and fabrication errors of the parabolic CRL [\cite{Lengeler1999}]. The modelling and functions presented here are based on the framework of physical optics [\cite{Paganin2006}] and are tailored to be used transparently with SRW [\cite{Chubar1998}], which already provides a model for the CRL [\cite{Baltser2011}]. This ideal model combines refraction and absorption for the stacked lenses. Optical imperfections from material inhomogeneities (voids, impurities) were later added to the CRL ideal model in SRW [\cite{Roth2014}]. 

We begin this proceeding by shortly revisiting the basic model of a transmission element in physical optics and that of an ideal CRL as presented in [\cite{Baltser2011}]. Expanding this model, we present the optical imperfections in refractive lenses in three different groups: \textit{i}-) misalignments of a single X-ray lens; \textit{ii}-) commonly encountered fabrication errors such transverse offsets as well as tilts of the individual parabolic sections; \textit{iii}-) and other sources of deviations from the parabolic shape modelled with either polynomial decomposition of error functions or by using metrology data. Each newly added feature is accompanied by a calculation of the residual thickness error, its impact on focusing by CRL and the beam caustic in the vicinity of the focal spot\footnote{All simulations shown throughout Figs.~\ref{fig:ideal_CRL}-\ref{fig:tilt_fs_CRL} and in Fig.~\ref{fig:Strehl} have similar conditions, that is, they model misalignments, fabrication errors or arbitrary residual errors to a single 2D-Beryllium lens with nominal radius $R=50~\mu\text{m}$, geometric aperture $A_{\diameter}=440~\mu\text{m}$ and $t_\text{wall}=20~\mu$m at 8~keV in fully-coherent simulations. The application of the presented modelling to partially-coherent simulations methods [\cite{Chubar2011}] is immediate.}. The code main functions implementing the ideal CRL and describing optical imperfections in refractive lenses are subsequently presented and discussed. We finish by discussing applications and future implementations. 

\section{X-RAY LENS MODELLING}\label{sec:lens_modelling} 
For a weak scatterer under the paraxial optics approximation, the transmission element in projection approximation in the physical optics framework is given by [\cite{Paganin2006}]:
\begin{align}\label{eq:transmission}
    \mathrm{T}\big[\Delta_z(x,y)]~\bullet &=\sqrt{\mathrm{T}_\text{BL}(x,y)}\exp{\big[i\phi(x,y)\big]}~\bullet,
\end{align}{}
with:
\begin{subequations}
\begin{align}   
    \mathrm{T}_{\text{BL}}(x,y)&=\exp{\big[-2k\beta(x,y)\Delta_z(x,y)\big]}\label{eq:aux_funcs_transa}  \\
    &=\exp{\big[-\mu(x,y)\Delta_z(x,y)\big]},\nonumber\\
    \phi(x,y)&=-k\delta(x,y)\Delta_z(x,y).\label{eq:aux_funcs_transb}
\end{align}
\end{subequations}
$k$ is the wavenumber, the index of refraction is written as $n=1-\delta+i\cdot\beta$, the projected thickness along the $z-$axis (beam propagation direction) is given by $\Delta_z$ and it depends on the transverse coordinates $(x,y)$. The $\bullet$ symbol represents the electric field to which the transmission operator is applied to. Eq.~\ref{eq:aux_funcs_transa} shows the absorption experienced by the wave-field when passing through matter (Beer-Lambert law) and Eq.~\ref{eq:aux_funcs_transb} shows the phase-shift. The coefficient multiplying $\Delta_z$ in $\mathrm{T}_{\text{BL}}$ is know as linear attenuation coefficient $\mu$.

\subsection{The ideal CRL}\label{sec:lens_ideal}

\begin{figure}[t]
    \centering
    {\includegraphics[width=0.8\linewidth]{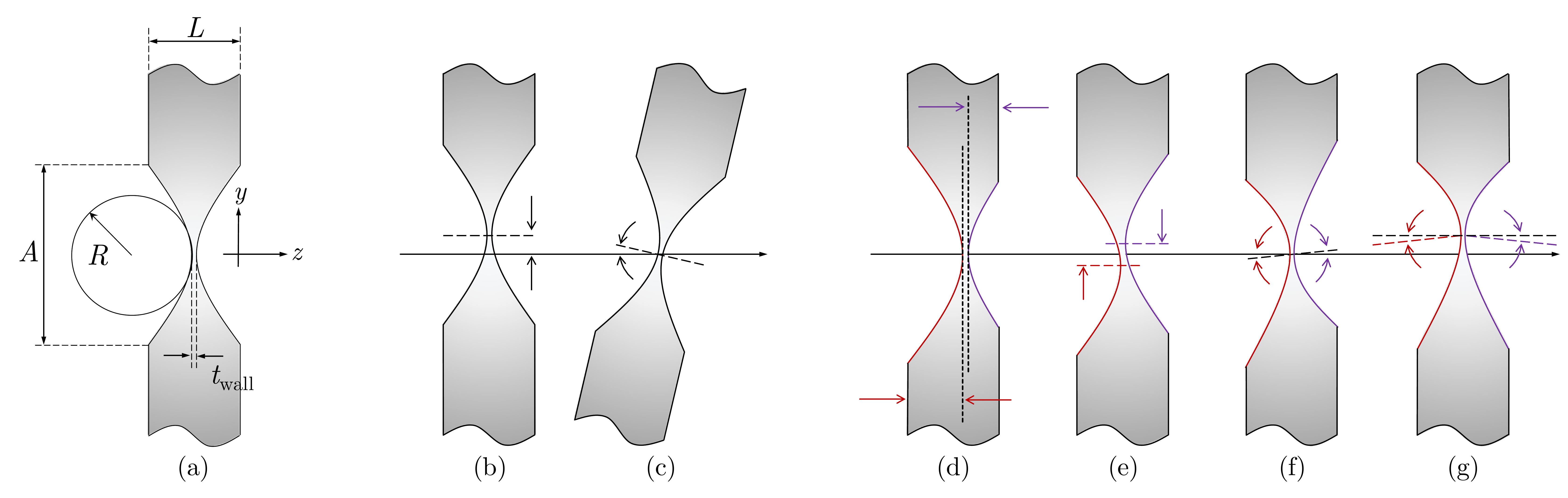}}
    \caption[]{\small (a) ideal lens for reference. Lens typical misalignments are the (b) transverse offset and the (c) tilt or a combination of both. Common fabrication errors include the (d) longitudinal offset of the parabolic section, (e) transverse offset of the parabolic section and (f)-(g) tilted parabolic sections.}
    \label{fig:lens_cuts}
\end{figure}

\begin{figure}[t]
    \centering
    {\includegraphics[width=.8\linewidth]{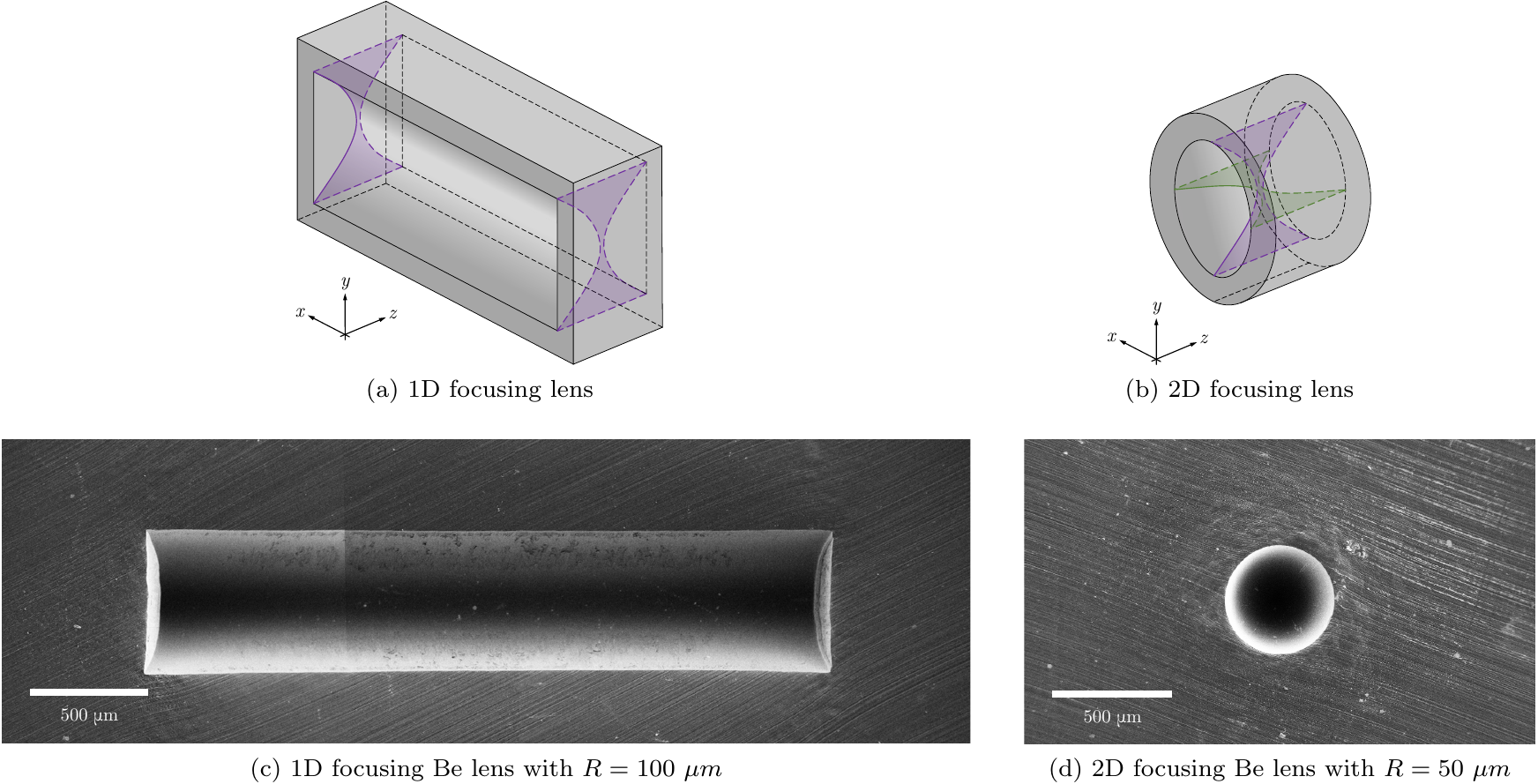}}
    \caption[1D and 2D focusing X-ray lenses]{\small 1D (left) and 2D focusing (right) X-ray lenses. The top row shows a 3D rendering of such lenses with emphasis on the parabolic profile - shaded in purple is the vertical profile and in green, the horizontal profile. Bottom row shows scanning electron microscope (SEM) images of two Be lenses. Due to the limited field of view, image (c) is stitched, which explains the colour discontinuation on the left side of the image.} 
    \label{fig:1D2D_lenses}
\end{figure}

At any point inside the geometric aperture $A$ of a single bi-concave paraboloidal X-ray lens, the projected thickness $\Delta_z$ can be calculated as:
\begin{equation}\label{eq:ProjecThick}
    \Delta_z(x,y) = 
     \begin{cases}
      \cfrac{x^2}{R_x}+\cfrac{y^2}{R_y}+\text{t}_\text{wall}, &\quad\forall~(x,y) \in A,\\
      L, &\quad\text{otherwise}.
     \end{cases}
\end{equation}
Here, $R_x$ and $R_y$ represent the apex radius of curvature in the horizontal and vertical directions, $\text{t}_\text{wall}$ is the distance between the apices of the parabolic surfaces and L is the lens total thickness as indicated by Fig.~\ref{fig:lens_cuts}(a). The lens geometric aperture is related to lens intrinsic parameters by:
\begin{equation}\label{eq:A}
    A = 2\sqrt{(L-\text{t}_\text{wall})R}.
\end{equation}{}
The focal length of such bi-concave lens is related to the curvature radius $R$ by:
\begin{equation}\label{eq:CRL_classic}
    f = \frac{R}{2\delta}.
\end{equation}{}
If the lens being modelled is a 1D focusing element, that is a cylinder with parabolic section, one of the radii goes to infinity to account for the non-curved surface. The geometric aperture in this direction is not given by Eq.~\ref{eq:A}, but arbitrarily chosen (cf. Fig.~\ref{fig:1D2D_lenses}). Eq.~\ref{eq:ProjecThick} can be substituted into Eqs.~\ref{eq:aux_funcs_transa} and \ref{eq:aux_funcs_transb} to retrieve the complex transmission element expression for an X-ray lens:
\begin{eqnarray}\label{eq:TE_singlelens}
    \mathrm{T}_{\text{single lens}}(\Delta_z)~\bullet =
    \exp{\bigg(-\frac{2\pi}{\lambda}\beta\Delta_z \bigg)}
    \times\exp{\bigg(-i\frac{2\pi}{\lambda}\delta\Delta_z \bigg)} \bullet.
\end{eqnarray}{}
Eq.~\ref{eq:TE_singlelens}, the single lens model, accounts for the absorption (first exponential) and phase shift (second exponential) from a single X-ray lenslet. Stacked lenses can be simulated either by a single-lens-equivalent representation or by using multi-slice-like techniques, where the X-ray beam is propagated using free-space propagators in between the optical elements [\cite{Celestre2020}].

\begin{figure}[h]
        \centering
        \subfloat[PSF]{\includegraphics[height=5cm]{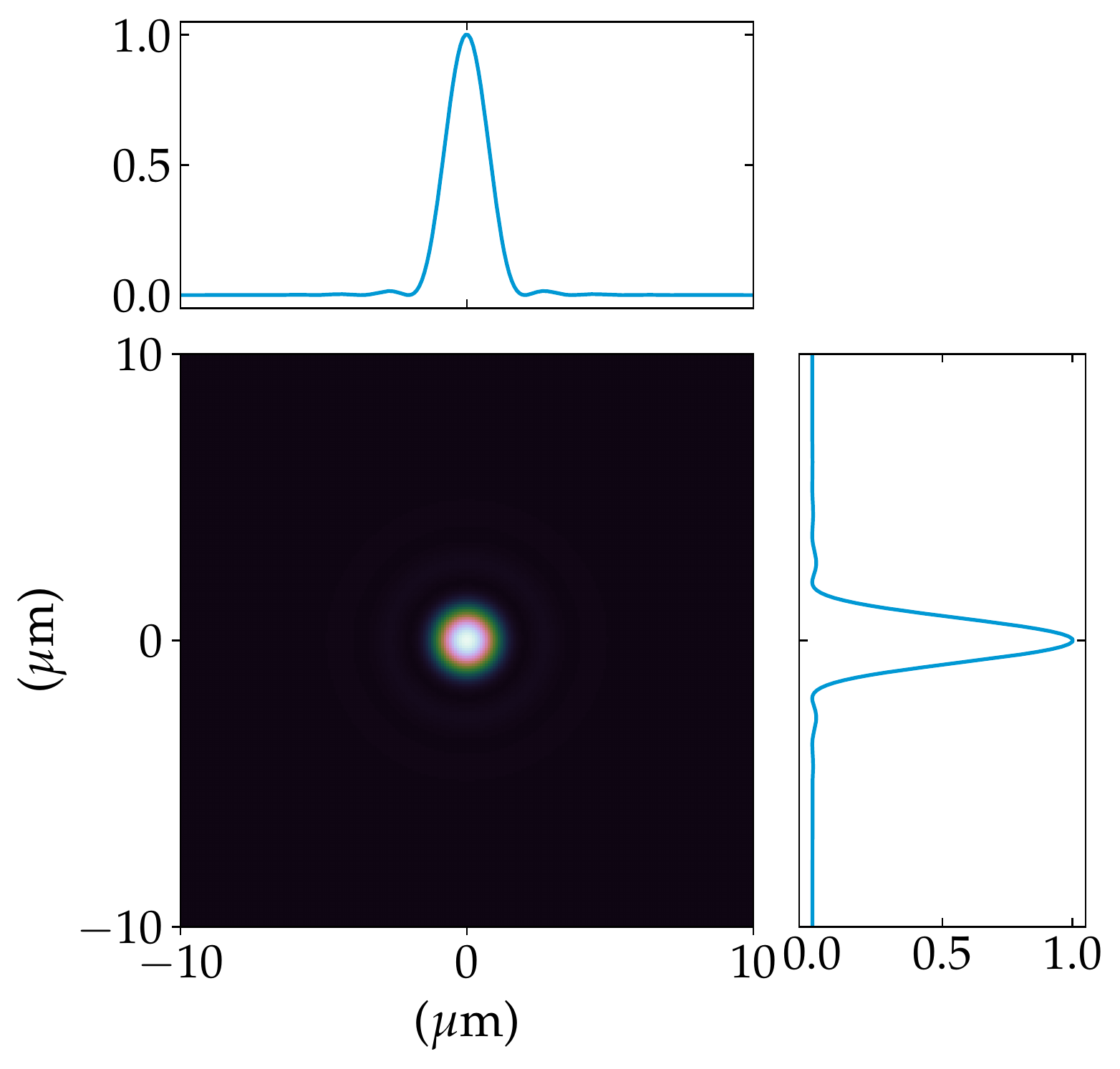}}\hspace{0.1cm}
        \subfloat[vertical caustics]{\includegraphics[height=3.5cm]{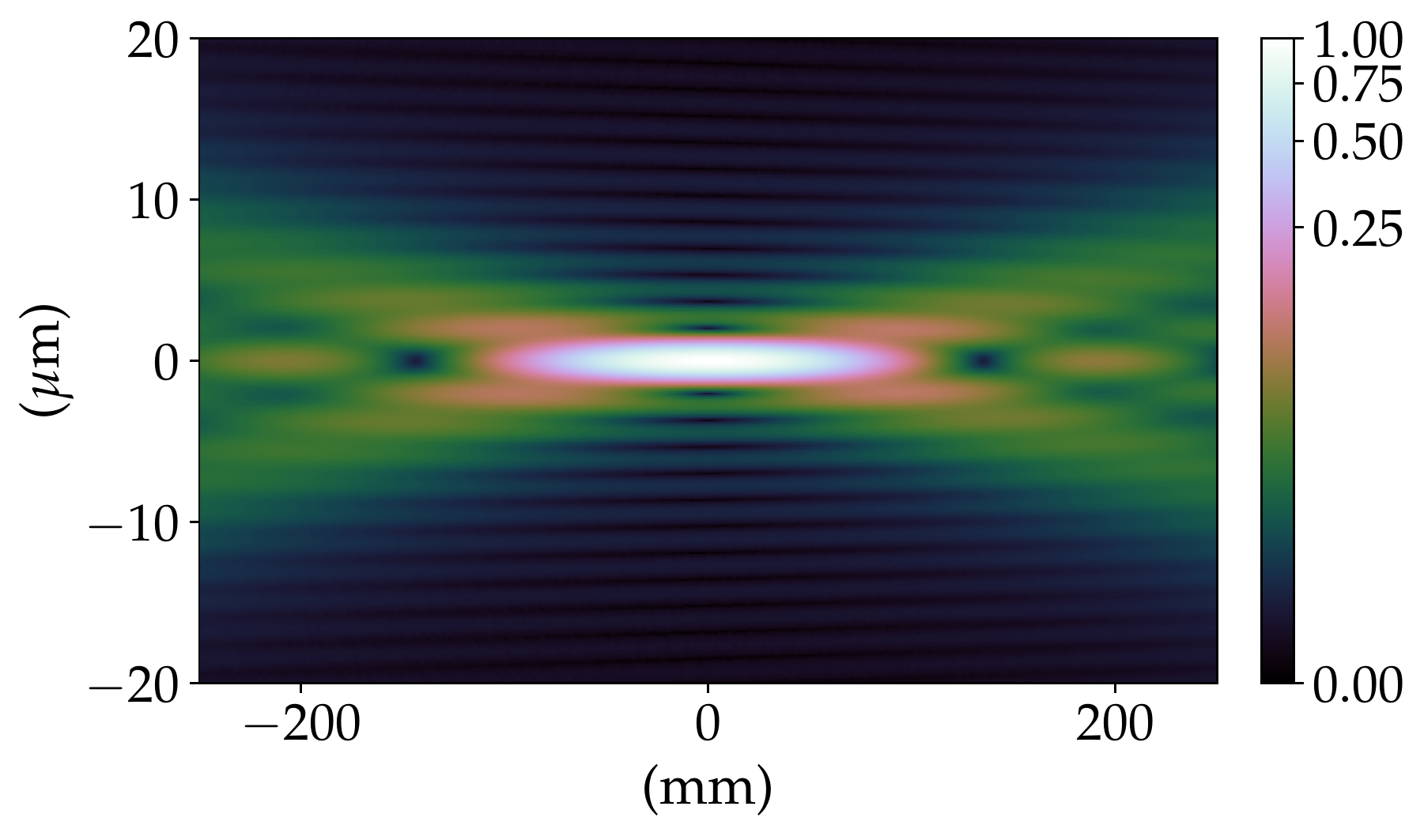}}
        \caption{\small Simulations of a single ideal 2D-Beryllium lens with nominal radius $R=50~\mu\text{m}$, geometric aperture $A_{\diameter}=440~\mu\text{m}$ and $t_\text{wall}=20~\mu$m at 8~keV  - cf. Fig~\ref{fig:lens_cuts}(a). (a) Point-spread function with cuts centred in $(0,0)$ and (b) the vertical beam caustics from -250~mm to 250~mm in respect to the focal plane.} \label{fig:ideal_CRL}
\end{figure}

\subsection{Describing optical imperfections in refractive lenses}\label{sec:lens_t_et_o}

In the paraxial approximation, the parabolic shape for a refracting surface is generally regarded as the ideal shape\footnote{The shape of a focusing refracting surface can be derived from the Fermat's principle, but the parabolic shape is generally regarded as a good approximation. Large apertures are often necessary when very small focused beams are required, but increasing the geometric aperture of the optical element causes the parabolic approximation to under-perform. Several aspheric surface shapes for different focusing conditions were reported in [\cite{SanchezdelRio2012}] - cf. Fig~4. For a deeper discussion on aspheric surfaces in the context of optics, please, refer to [\cite{Schulz1988}].} for minimising aberrations. It is legit, then, to define as errors any deviation from this ideal parabolic form\footnote{Such definition, however, leaves out discrepancies in the radius of curvature $R$ (designed vs. \textit{de facto}) and the associated defocus it may cause. Discrepancies between designed and executed lenses may render them to be labelled as out-of-specification and may cause the system to under-perform, but are not deviations of the parabolic shape, provided the ideal parabolic shape takes into account the \textit{de facto} radius of curvature. Accounting for such discrepancies can be done using the ideal model described by the transmission element $\mathrm{T}_{\text{single lens}}(\Delta_z)$ (cf. Eq.~\ref{eq:TE_singlelens}) using the \textit{de facto} radius of curvature.} regardless of its origin. The phase errors induced by an ideal lens misalignment will be presented first, then the typical fabrication errors of bi-concave lenses will be presented shortly after. The misalignment and fabrications errors presented in this section were derived from the accumulated experience in handling beryllium and aluminium bi-concave embossed lenses, which are the most available throughout beamlines in diverse synchrotron facilities. However, the modelling presented here is generic and can be applied to a wide-range of CRL from diverse fabrication processes\footnote{cf. Table~1 from the supplementary material relative to [\cite{Roth2017}].}. Fig.~\ref{fig:ideal_CRL} shows the focusing performance of a single ideal 2D-Beryllium lens with nominal radius $R=50~\mu\text{m}$, geometric aperture $A_{\diameter}=440~\mu\text{m}$ and $t_\text{wall}=20~\mu$m at 8~keV using the SRW basic modelling. 

\subsection{Misalignments}\label{sec:misalignments}

Misalignments of optical systems are not optical errors \textit{per se} as they can be mitigated by ensuring proper alignment is done; they will, however, cause changes to the ideal parabolic phase profile if left uncorrected and will affect the optical performance of the system. Although aligning a CRL stack is possible\footnote{The possibility of realignment of the CRL depends on where and how they are installed in the beamline. If their installation is on a bulky transfocator [\cite{Vaughan2011}], their realignment is more difficult to be performed. However, when used as a final focusing element, enclosed in small casings or compact transfocators - cf. Fig.~3 in [\cite{Lengeler1999}] and [\cite{Kornemann2017, Narikovich2019}], their realignment can be done more easily.}, the individual lenslets usually cannot be aligned to each other, hence the interest in modelling such misalignments.

\subsubsection*{Transverse offset}

\begin{figure}[t]
        \centering
        \subfloat[residues]{\includegraphics[height=3.cm]{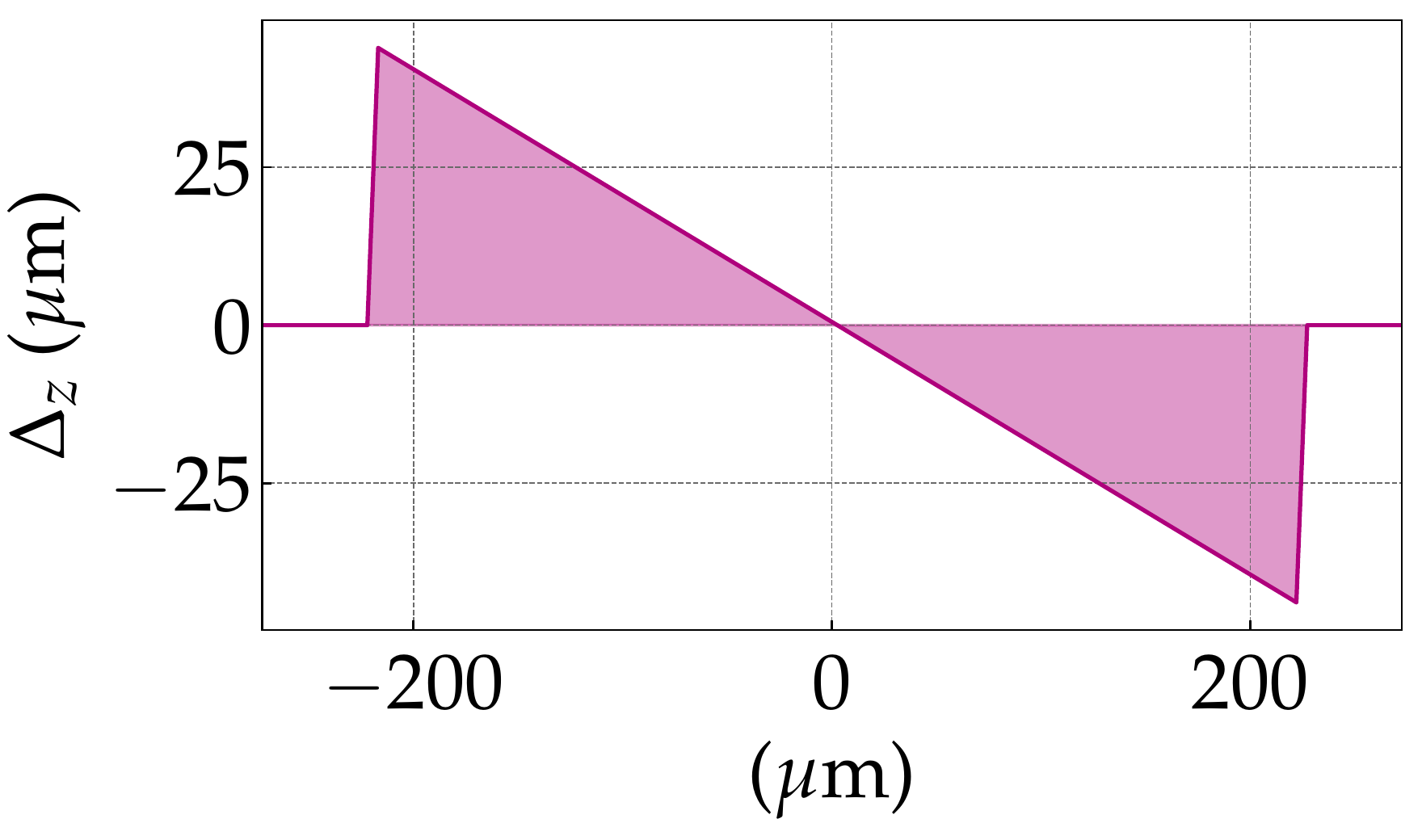}}\hspace{0.1cm}
        \subfloat[PSF]{\includegraphics[height=5cm]{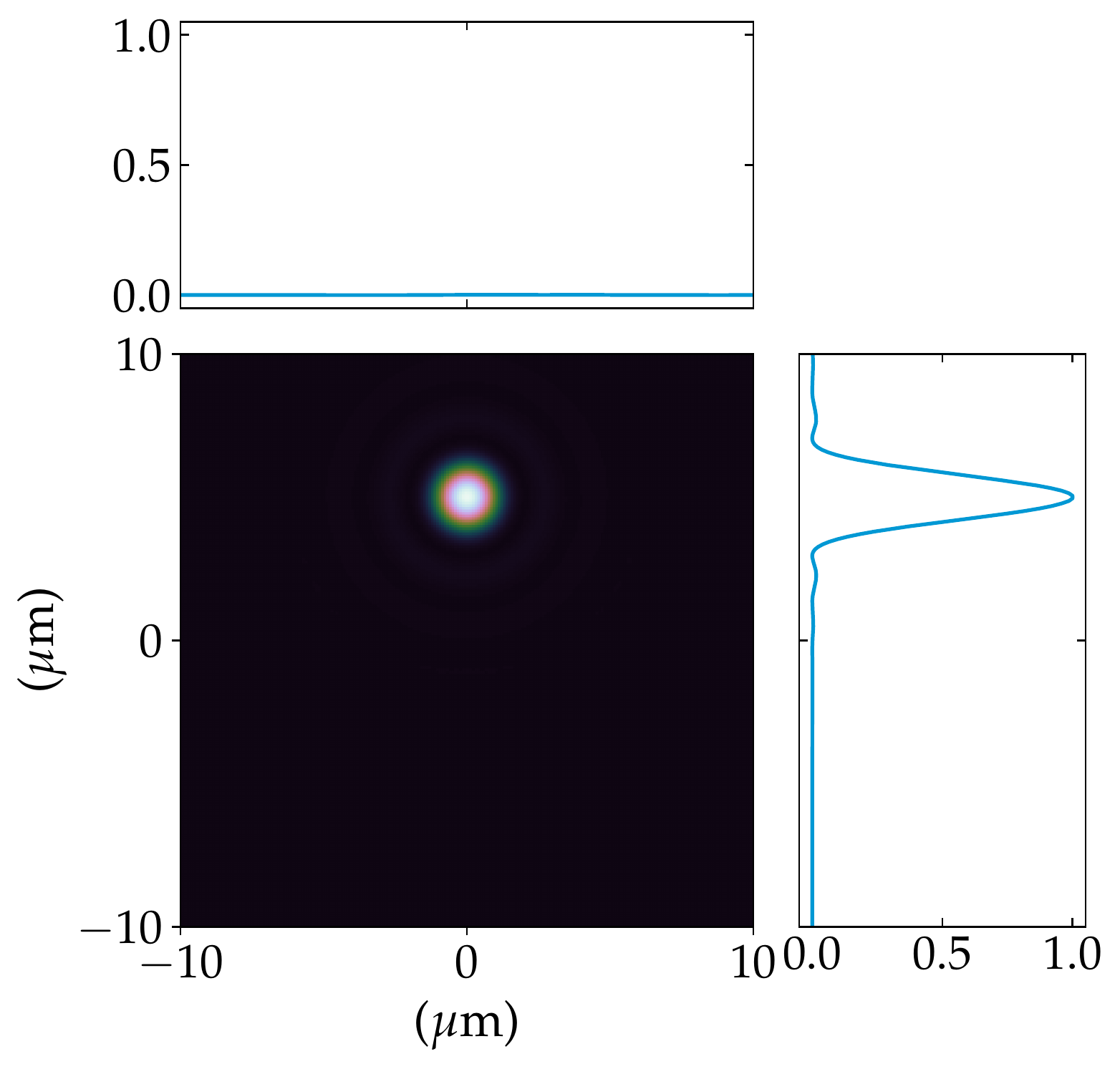}}\hspace{0.1cm}
        \subfloat[vertical caustics]{\includegraphics[height=3.5cm]{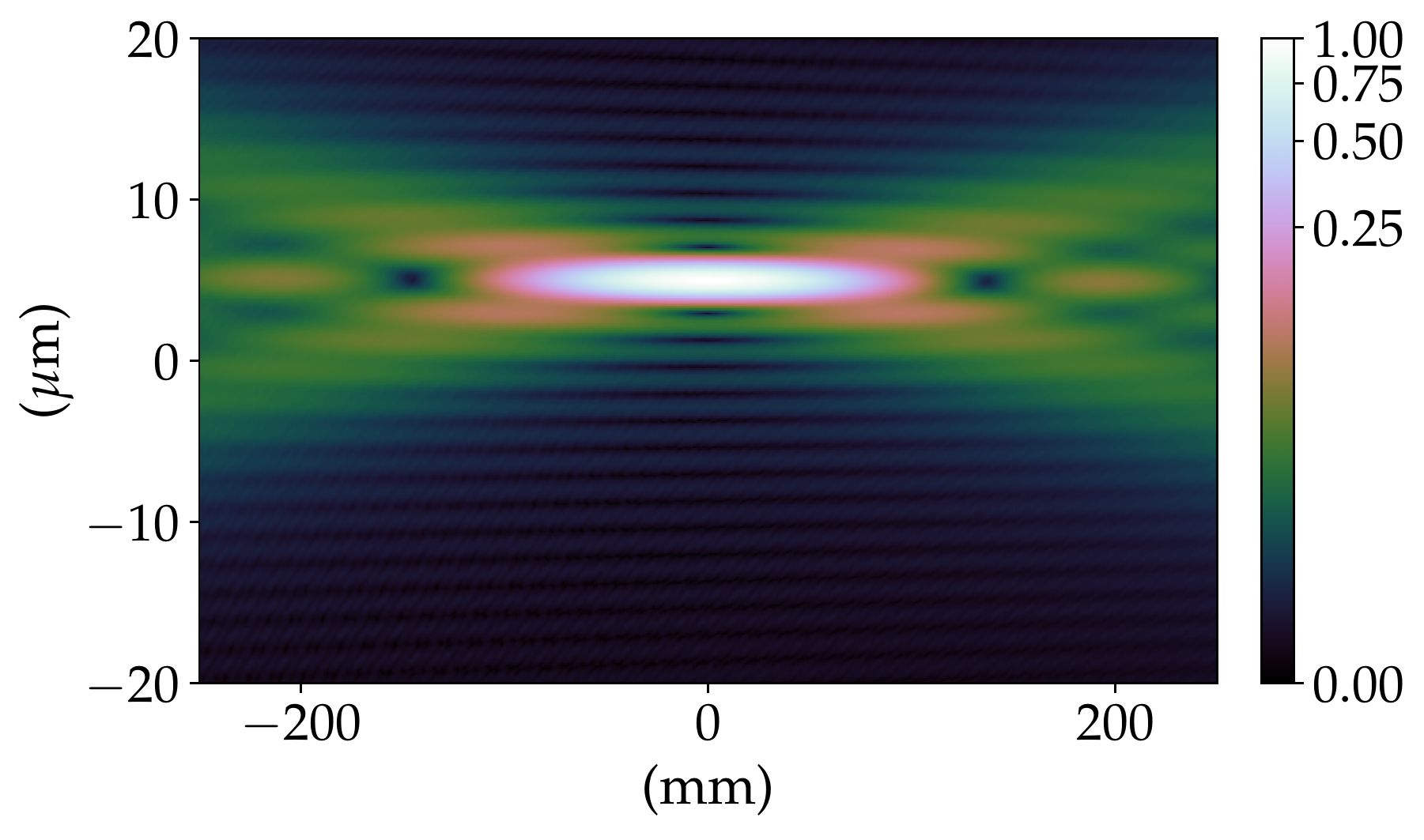}}
        \caption{\small Simulations of an ideal lens shifted by $\Delta_y=5~\mu$m - cf. Fig~\ref{fig:lens_cuts}(b). (a) Residual thickness, (b) point spread function with cuts centred in $(0,0)$ and (c) the vertical beam caustics from -250~mm to 250~mm in respect to the focal plane. 2D-Beryllium lens with nominal radius $R=50~\mu\text{m}$, geometric aperture $A_{\diameter}=440~\mu\text{m}$ and $t_\text{wall}=20~\mu$m at 8~keV.} \label{fig:shifted_CRL}
\end{figure}

Shifting transversely a single element an transverse distance $(\Delta_x,\Delta_y)$ can be simply done by calculating $\Delta_z(x-\Delta_x,y-\Delta_y)$ in Eq.~\ref{eq:ProjecThick}. The shifted element is depicted in Fig.~\ref{fig:lens_cuts}(b). For a pair of coordinates $(x,y)$:
\begin{equation}\label{eq:ProjecThick_misaligned}
    \Delta_z(x-\Delta_x,y-\Delta_y) = 
        \begin{cases}
      \cfrac{(x-\Delta_x)^2}{R_x}+\cfrac{(y-\Delta_y)^2}{R_y}+\text{t}_\text{wall}, &\quad\forall~(x-\Delta_x,y-\Delta_y) \in A,\\
      L, &\quad\text{otherwise}.
        \end{cases}
\end{equation}   
Eq.~\ref{eq:ProjecThick_misaligned} is the ideal parabolic profile of a bi-concave lens given by Eq.~\ref{eq:ProjecThick} with its vertices centred around $(\Delta_x,\Delta_y)$. While a single transversely shifted lens considered on its own is innocuous, piling up several shifted lenses has impacts on the overall accumulated phase parabolic shape and resulting geometric aperture. Although the exact effect of relative misaligments between individual lenses on the phase of the wave-field depends on the distance between lenslets, the energy, footprint and divergence of the X-ray beam, some insight can be gained by considering the individual focusing elements as thin-optical elements in intimate contact. Consider $N$ stacked lenses transversely misaligned with their transverse distance to the optical axis given by $(\Delta_{x_j},\Delta_{y_j})$, with $j=1,~2,~...,~N$. Within the intersection of their geometric apertures, the accumulated thickness is given by:
\begin{align}\label{eq:ProjecThick_Nmisaligned}
    \Delta_{z_\Sigma}(x,y) &= \sum\limits_{j=1}^N \Delta_z(\Delta_{x_j},\Delta_{y_j})\nonumber\\
    &=\sum\limits_{j=1}^N \underbrace{\frac{x^2}{R_{x_j}}+\frac{y^2}{R_{y_j}}}_\text{(I)}
    -\underbrace{2x\frac{\Delta_{x_j}}{R_{x_j}} - 2y\frac{\Delta_{y_j}}{R_{y_j}}}_\text{(II)}
    +\underbrace{\frac{\Delta_{x_j}^2}{R_{x_j}}+\frac{\Delta_{y_j}^2}{R_{y_j}}+\text{t}_{\text{wall}_j}}_\text{(III)}.
\end{align}
The first term in Eq.~\ref{eq:ProjecThick_Nmisaligned}, ($\text{I}$) is a quadratic term and it indicates ideal focusing as in Eq.~\ref{eq:ProjecThick}. The residual terms ($\text{II}$) and ($\text{III}$) are a linear term in $x$ and $y$ and a constant offset term, respectively. The projected thickness and the phase are linearly proportional, so the residual accumulated thickness translates directly into residual accumulated phase and both terms can be used interchangeably - cf. Eq.\ref{eq:aux_funcs_transb}. The first residual term, i.e. ($\text{II}$), adds a linear phase to the wave-front and acts like a prism, not deforming the monochromatic wave-field, but redirecting it. At the focal plane, the image position is transversely shifted but no change to the intensity and phase profiles is added. Symmetrically shifted lenses\footnote{That is $\Delta_{x_m}=-\Delta_{x_n}$ or $\Delta_{y_m}=-\Delta_{y_n}$ for $m,n\in(1,~2,~...,~N)$.} make ($\text{II}$) goes to zero. The residual terms in ($\text{III}$) add a constant phase offset to the transmitted beam.  The effects of the transverse offset to a single X-ray lens are shown in Fig.~\ref{fig:shifted_CRL}.

\subsubsection*{Tilted lens}
\begin{figure}[t]
        \centering
        \subfloat[residues]{\includegraphics[height=3cm]{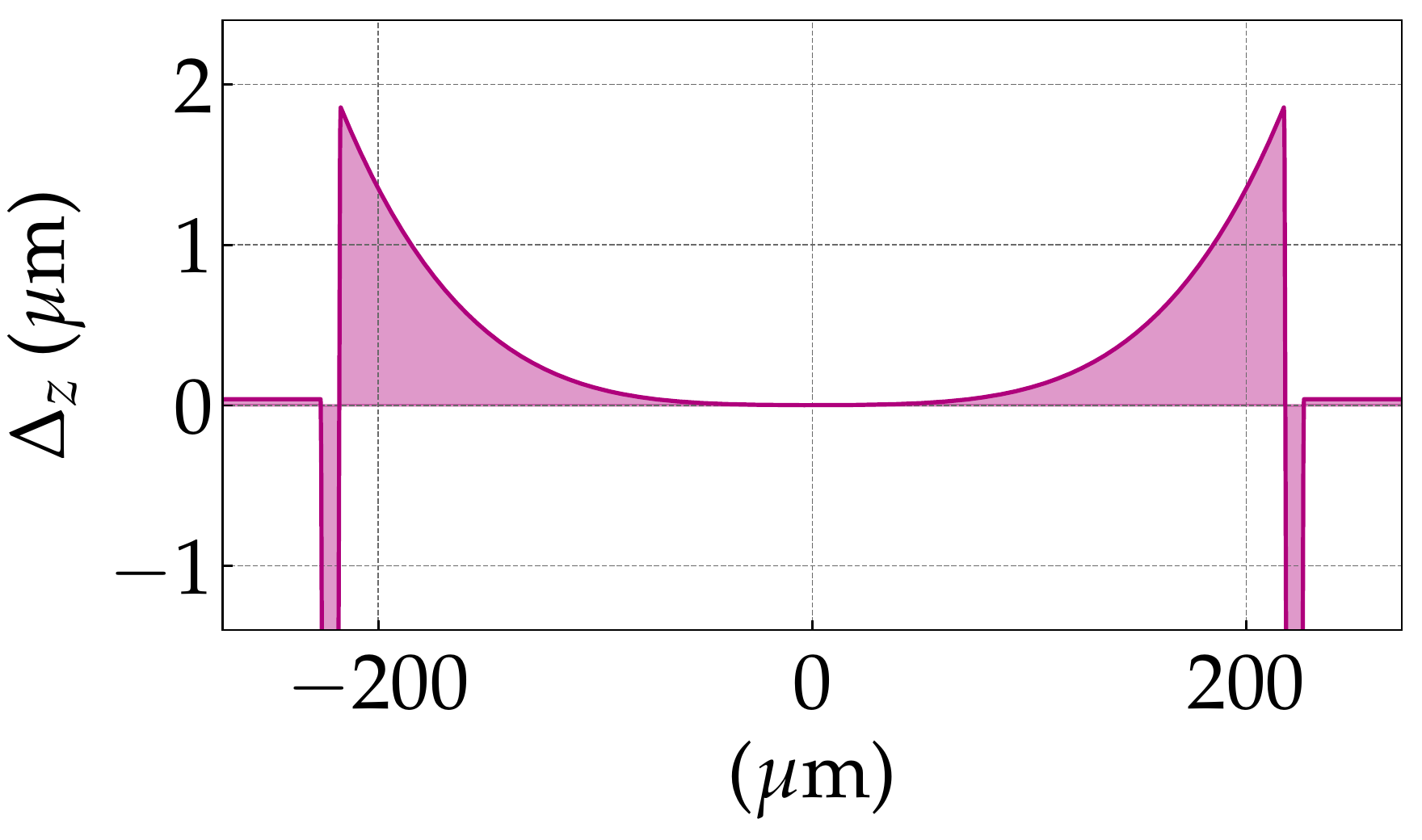}}\hspace{0.1cm}
        \subfloat[PSF]{\includegraphics[height=5cm]{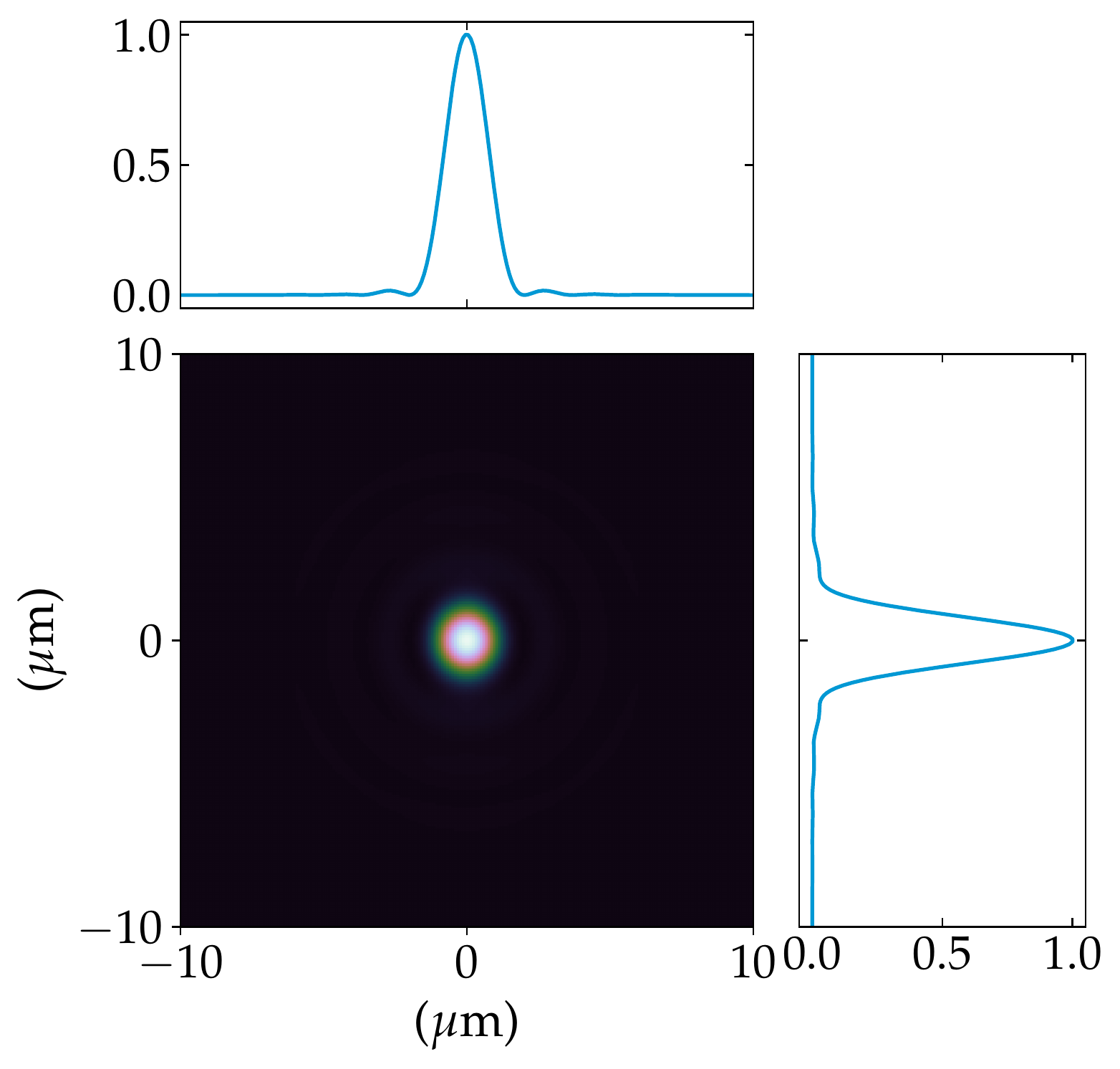}}\hspace{0.1cm}
        \subfloat[vertical caustics]{\includegraphics[height=3.5cm]{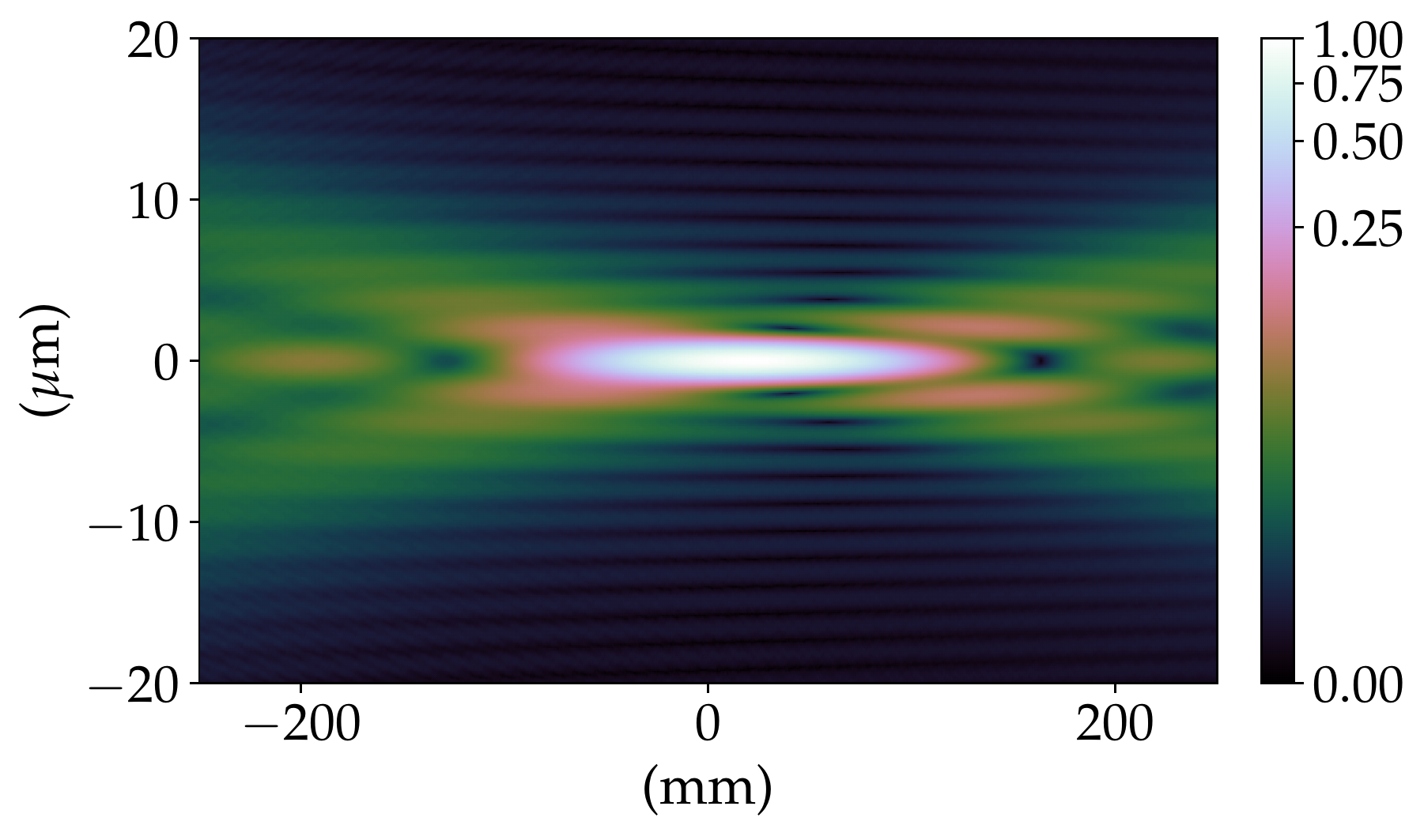}}
        \caption{\small Simulations of an ideal lens tilted by $\theta_x=1^{\circ}$ - cf. Fig~\ref{fig:lens_cuts}(c). (a) Residual thickness, (b) point spread function with cuts centred in $(0,0)$ and (c) the vertical beam caustics from -250~mm to 250~mm in respect to the focal plane. 2D-Beryllium lens with nominal radius $R=50~\mu\text{m}$, geometric aperture $A_{\diameter}=440~\mu\text{m}$ and $t_\text{wall}=20~\mu$m at 8~keV. The profile shown in (a) is proportional to the 4$^{\text{th}}$ power of the lateral coordinates in the direction of the shift. This contributes for the elongation of the beam along the propagation direction and shift on the focal plane position as evidenced by (c). This is a typical sign of spherical aberrations.} \label{fig:tilted_CRL}
\end{figure}

When rotating a lens in space as shown in Fig.~\ref{fig:lens_cuts}(c) and calculating its projected thickness, it is helpful to decouple the rotation of the front and back surfaces. This can be done by defining a point cloud in Cartesian coordinates:
\begin{subequations}\label{eq:point_cloud}
    \begin{align}
    z_\text{front surface}(x-\Delta_x,y-\Delta_y) &=\frac{\Delta_z(x-\Delta_x,y-\Delta_y)}{2},\\
    z_\text{back surface}(x-\Delta_x,y-\Delta_y) &= -\frac{\Delta_z(x-\Delta_x,y-\Delta_y)}{2}
    \end{align}
\end{subequations}{}
where $\Delta_z(x-\Delta_x,y-\Delta_y)$ is given by Eq.~\ref{eq:ProjecThick_misaligned}. The projected thickness is given by:
\begin{align}\label{eq:point_cloud_thickness}
    \Delta_z(x,y) = z_\text{front surface}(x,y) - z_\text{back surface}(x,y),
\end{align}{}
provided those are calculated on the same grid $(x,y)$. A tilted lens can be described by rotation matrices in three dimensions. The transformation matrices allowing a rotation $\theta_{x,y,z}$ around each of the Cartesian axis are~[\cite{House2016}]:
\begin{subequations}\label{eq:affine}
    \begin{align}
        R_x &= \begin{bmatrix}
                            1 & 0 & 0 &0\\
                            0 & \cos\theta_x & -\sin{\theta_x}  &0\\
                            0 & \sin\theta_x & \cos\theta_x &0\\
                            0 & 0 & 0 &1
            \end{bmatrix}  &&(\small{\text{rotation around the $x-$axis}}),\\
            R_y &= \begin{bmatrix}
                            \cos\theta_y & 0 & \sin\theta_y &0\\
                            0 & 1 & 0 &0\\
                            -\sin\theta_y & 0 & \cos\theta_y &0\\
                            0 & 0 & 0 &1
            \end{bmatrix}  &&(\small{\text{rotation around the $y-$axis}}),\\
            R_z &= \begin{bmatrix}
                            \cos\theta_z & -\sin\theta_z & 0 &0\\
                            \sin\theta_z & \cos\theta_z & 0 &0\\
                            0 & 0 & 1 &0\\
                            0 & 0 & 0 &1
            \end{bmatrix}  &&(\small{\text{rotation around the $z-$axis}}).
    \end{align}
\end{subequations}{}
Matrix multiplication is associative, which implies that if multiple rotations are involved, that is $R_x, R_y$ and $R_z$ are applied to a set of points $(x,y,z)$, an equivalent rotation matrix $R_{\theta}=R_zR_yR_x$ can be calculated and then, applied to those points is space: 
\begin{align}\label{eq:affine2}
    [x_\theta,y_\theta,z_\theta,1]^\text{T} & = R_zR_yR_x[x,y,z,1]^\text{T},\nonumber \\
     & = R_\theta[x,y,z,1]^\text{T},
\end{align}{}
where $(x_\theta,y_\theta,z_\theta)$ are the transformed $(x,y,z)$ coordinates after the $R_{\theta}=R_zR_yR_x$ rotation and the $^\text{T}$ in Eq.~\ref{eq:affine2} represents transposed matrices. The rotations given by $R_{\theta}$ have to be applied to both the front and back surfaces of the lens independently, with their respective point clouds given by Eqs.~\ref{eq:point_cloud}. In order to calculate the projected thickness along the optical axis, the rotated front and back surfaces have to be recalculated on a common grid, which is done by two-dimensional interpolation of $(x_\theta,y_\theta,z_\theta)$ to the original $(x,y)$ grid. The associative property allows for considerable computation time reduction, as the rotation can be done applying a single equivalent rotation matrix as opposed to three individual rotations. On the other hand, matrix multiplication is not commutative and the order of operations matter and should be specified\footnote{The implementation of the affine transformations for rotating CRLs in space follow the order: rotation around the $x-$axis ($R_x$), rotation around the $y-$axis ($R_y$) and then, rotation around the $z-$axis ($R_z$).}$^{,}$\footnote{For a deeper discussion on the properties of the affine transformations, coordinates systems and quaternions, please refer to appendices \textit{C}-\textit{E} in [\cite{House2016}].} when rotating a point cloud. Equations~\ref{eq:affine} have their pivot point centred in the origin of their axis, that is, around $(x,y,z)=(0,0,0)$ in Cartesian coordinates. It is possible to define arbitrary pivot points with a combination of translations and rotations. Tilting an optical element in space will introduce aberrations to the beam propagation and its focusing\footnote{The interest in tilted optical elements and compensation is an active field, as evidenced by the literature on that subject - cf. [\cite{Guizar-Sicairos2011,Zhou2019,Ali2020}].}. This is evidenced by the residual accumulated thickness in projection approximation shown in Fig.~\ref{fig:tilted_CRL}. It is also possible to see how the two surfaces (back and front) do not overlap, causing a slight reduction in the geometric aperture area, which is evidenced by the discontinuities in Fig.~\ref{fig:tilted_CRL}(a).

\subsection{Fabrication errors}\label{sec:fabrication}

Modelling the typical misalignment of X-ray lenses implies calculating the lateral displacements and rotations in space of an ideal X-ray lens. However, bi-concave lenses may also present misalignments between the front and back focusing surfaces, these are closely related to the manufacturing processes involved in the lens production. Here, the front and back focusing surfaces are treated independently, allowing to model longitudinal and transverse misalignments as well as tilts of the front and back focusing surface concerning the optical axis.

\subsubsection*{Longitudinal offset of the parabolic section}
Longitudinal offsets of the parabolic portions of a bi-concave X-ray lens appear when, for the same radius of curvature $R$, one parabolic portion is deeper than the other one - cf. Fig.~\ref{fig:lens_cuts}(d). The first eminent observation if that front and back surfaces will have different geometric apertures along the focusing direction. The new geometric aperture of the longitudinally offset parabolic profile can be calculated as:
\begin{equation}\label{eq:A_2}
    A_{\text{offset}} = 2\sqrt{[L-(\text{t}_\text{wall}+2\cdot\text{offset})]R},
\end{equation}{}
where a positive offset increases the apparent web thickness of the half lens to $\text{t}_\text{wall}/2+\text{offset}$ and decreases the geometric aperture for a fixed lens thickness. The aperture given by $ A_{\text{offset}}$ and the apparent web thickness are used in Eq.~\ref{eq:point_cloud_thickness} (cf. Eqs.~\ref{eq:ProjecThick_misaligned} and \ref{eq:point_cloud}) when calculating the projected thickness.
Longitudinal offsets do not affect the parabolic accumulated shape of a single lens and, consequently, do not impose any optical imperfection to an optical system based on such lenses. However, they are often encountered in real lenses\footnote{Specially in embossed lenses, where different penetration depths of the punches often lead to asymmetric lenses.} and merit the implementation in the lens modelling.

\subsubsection*{Transverse offset of the parabolic section}

\begin{figure}[t]
        \centering
        \subfloat[residues]{\includegraphics[height=3cm]{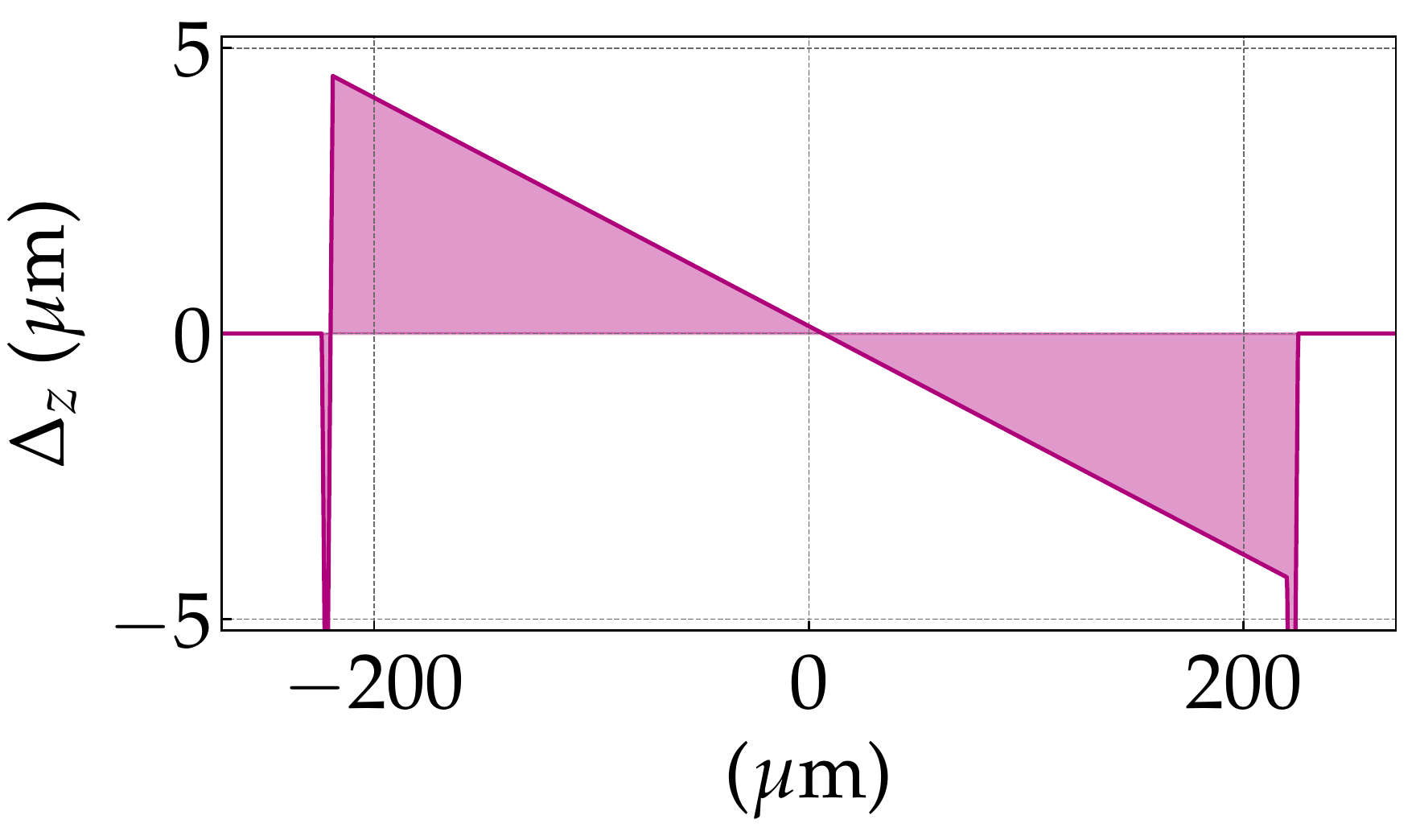}}\hspace{0.1cm}
        \subfloat[PSF]{\includegraphics[height=5cm]{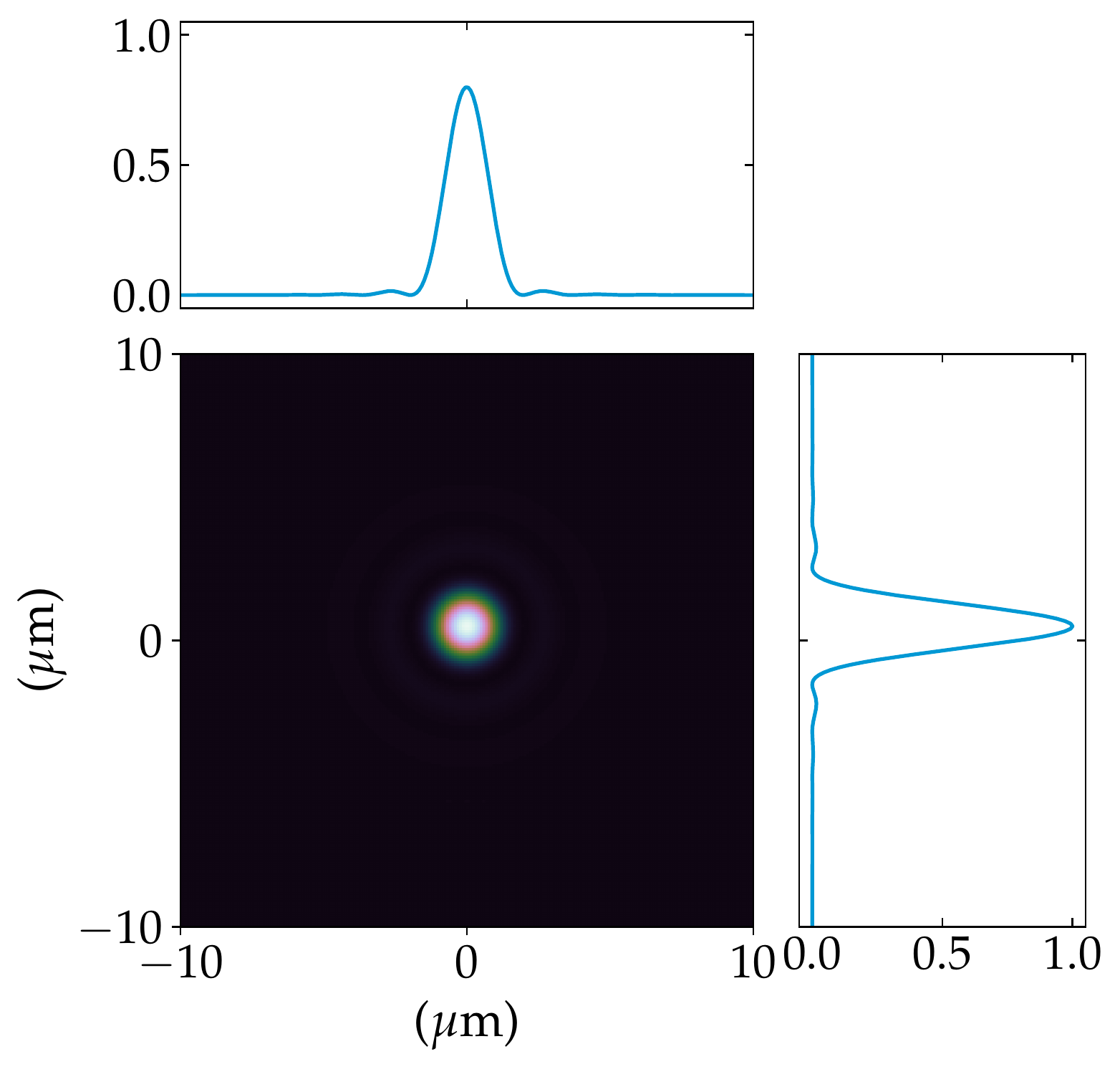}}\hspace{0.1cm}
        \subfloat[vertical caustics]{\includegraphics[height=3.5cm]{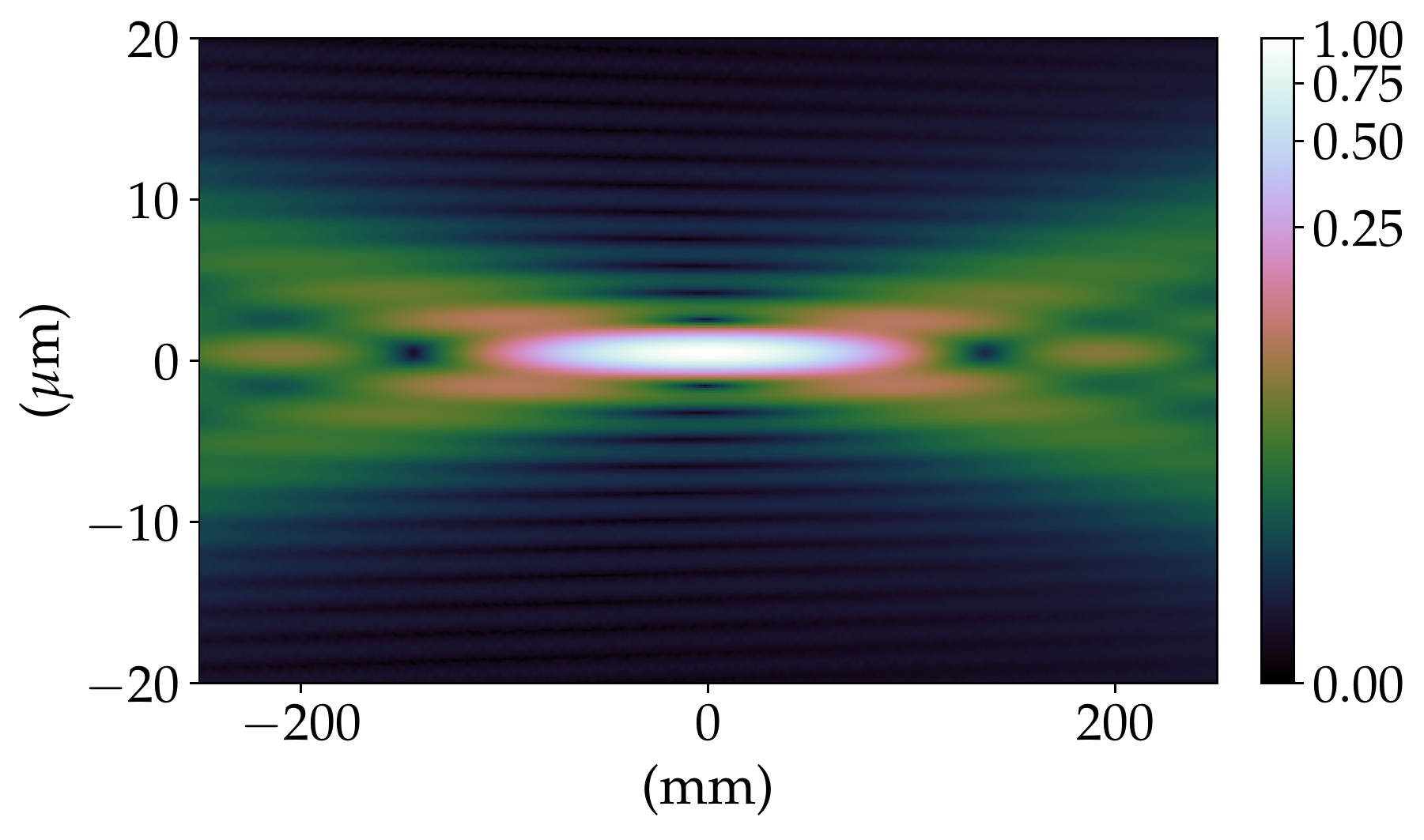}}
        \caption{\small Simulations of a lens with front focusing parabolic section shifted by $\Delta_y=-2~\mu$m and back focusing shifted by $\Delta_y=-3~\mu$m - cf. Fig~\ref{fig:lens_cuts}(e). (a) Residual thickness, (b) point spread function with cuts centred in $(0,0)$ and (c) the vertical beam caustics from -250~mm to 250~mm in respect to the focal plane. 2D-Beryllium lens with nominal radius $R=50~\mu\text{m}$, geometric aperture $A_{\diameter}=440~\mu\text{m}$ and $t_\text{wall}=20~\mu$m at 8~keV.} \label{fig:offset_fs_CRL}
\end{figure}

Although parallel to the optical axis, it is possible that the parabolic surfaces axes are not collinear. This is shown in Fig.~\ref{fig:lens_cuts}(e). The modelling of the transverse offset of the front or/and back surfaces of a lens concerning the optical axis is done by calculating the net offset of each surface, that is the sum of lens transverse offset with the front or/and back surface transverse offset, and applying it to Eqs.~\ref{eq:point_cloud} when calculating Eq.~\ref{eq:point_cloud_thickness}. The effects on the residual accumulated phase of non-collinear parabolic surfaces are the same as the one described in the section \textit{Transverse offset} from \S\ref{sec:misalignments}~-~\textit{\nameref{sec:misalignments}}, that is, the presence in the residual phase of a linear and a constant term, which can be seen in Fig.~\ref{fig:offset_fs_CRL}.

\subsubsection*{Tilted parabolic section}

\begin{figure}[t]
        \centering
        \subfloat[residues]{\includegraphics[height=3cm]{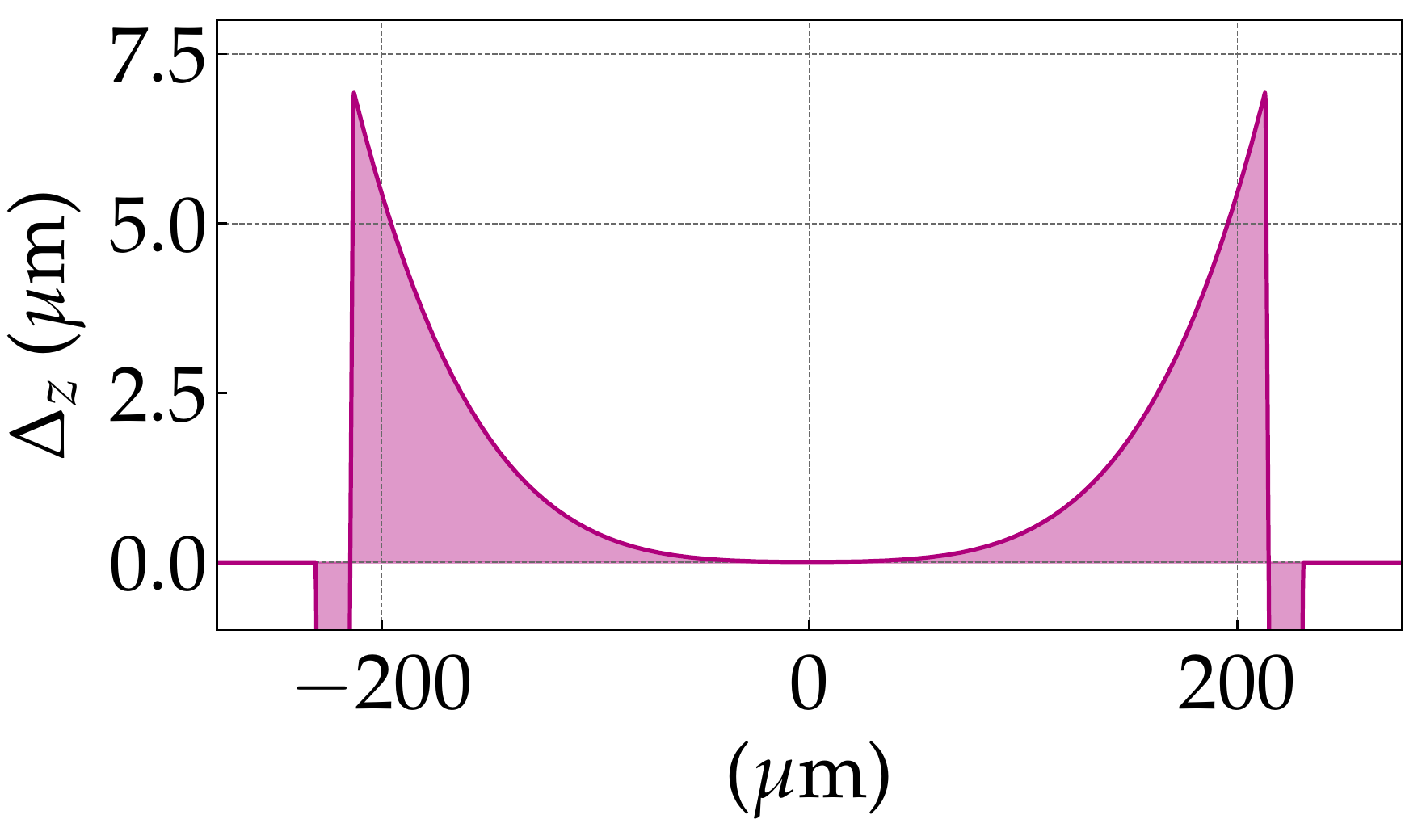}}\hspace{0.1cm}
        \subfloat[PSF]{\includegraphics[height=5cm]{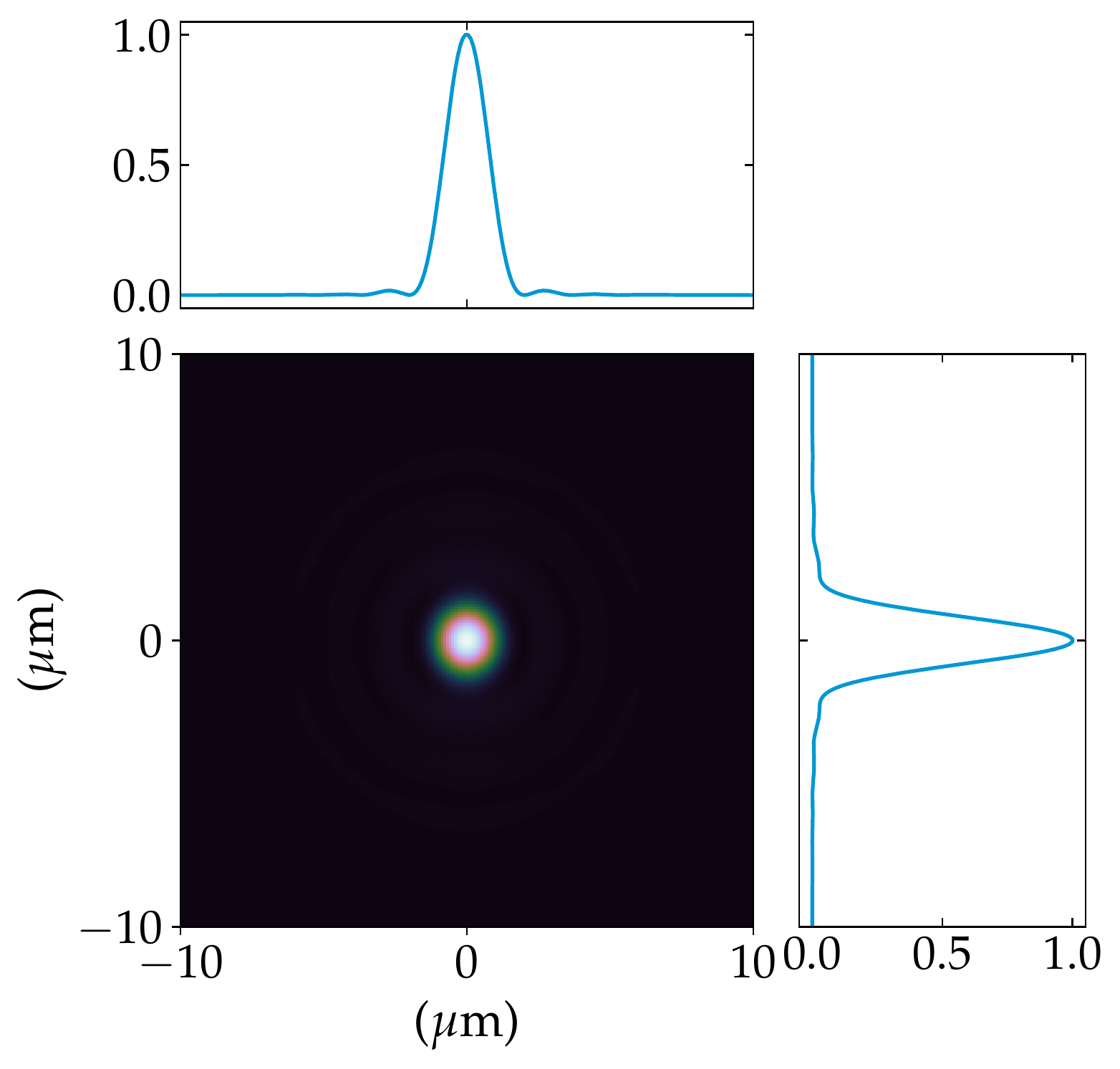}}\hspace{0.1cm}
        \subfloat[vertical caustics]{\includegraphics[height=3.5cm]{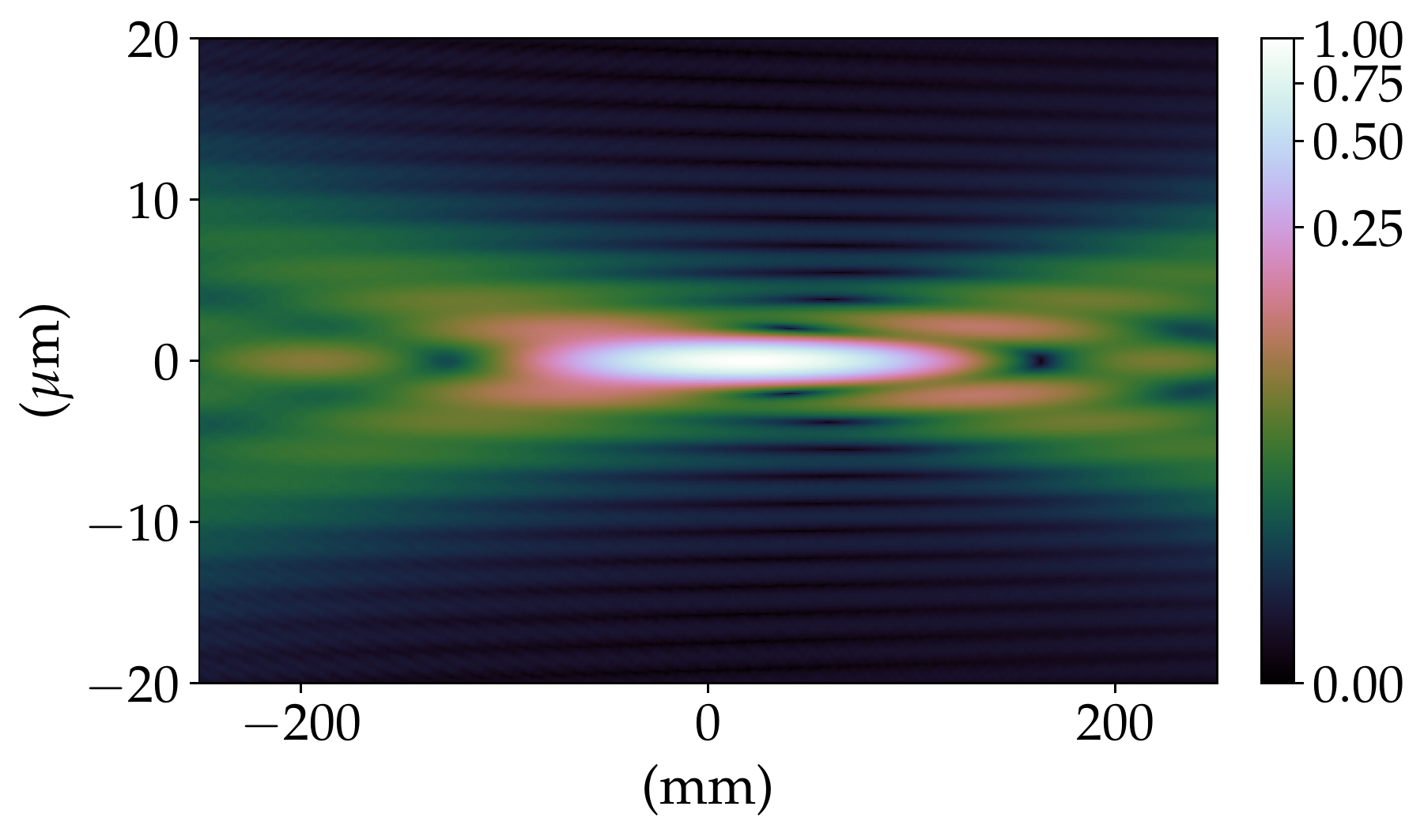}}\\
        \caption*{Symmetric tilt of front- and back- surfaces by $\theta_{x_{\text{front}}}=\theta_{x_{\text{back}}}=1^{\circ}$}
        \subfloat[residues]{\includegraphics[height=3cm]{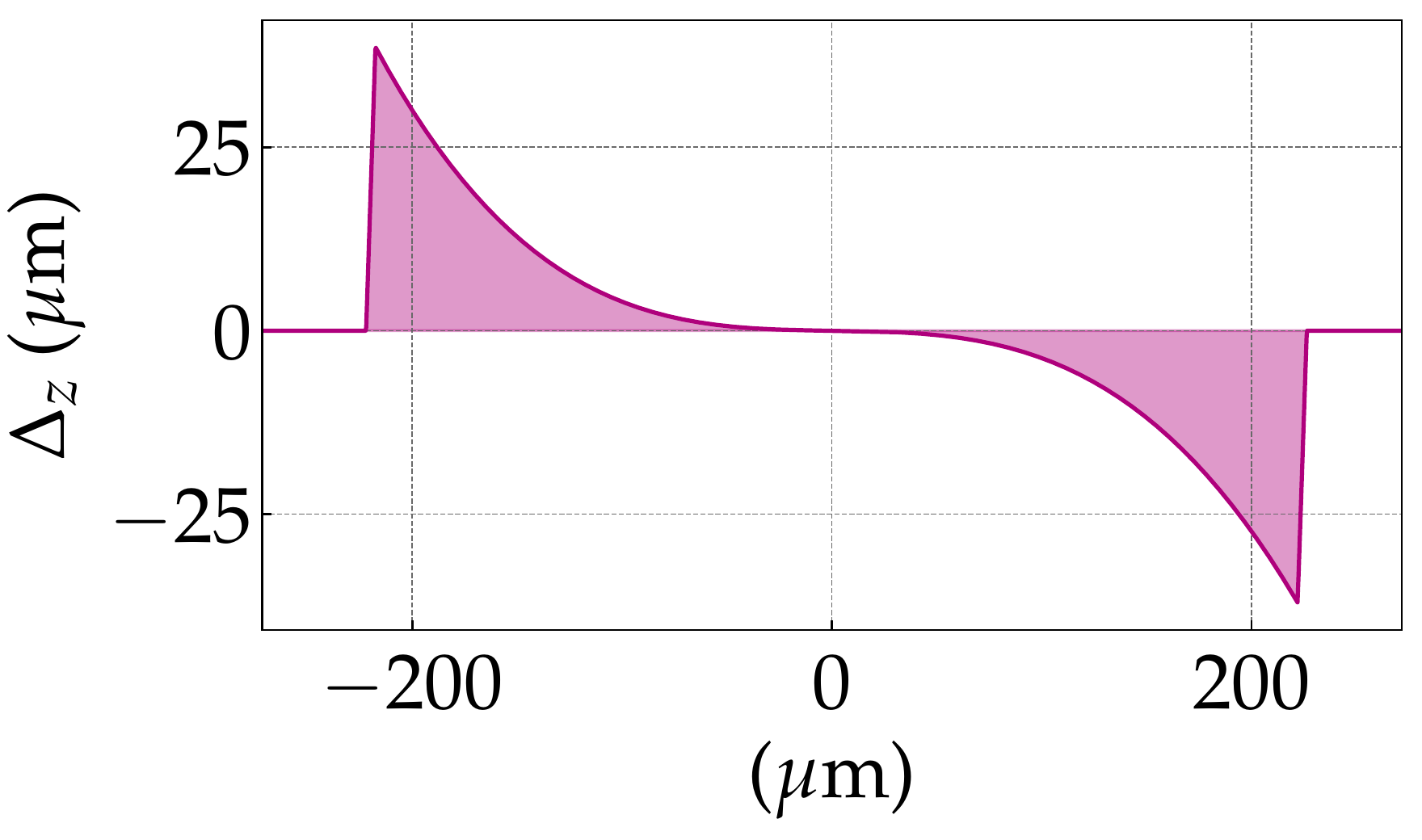}}\hspace{0.1cm}
        \subfloat[PSF]{\includegraphics[height=5cm]{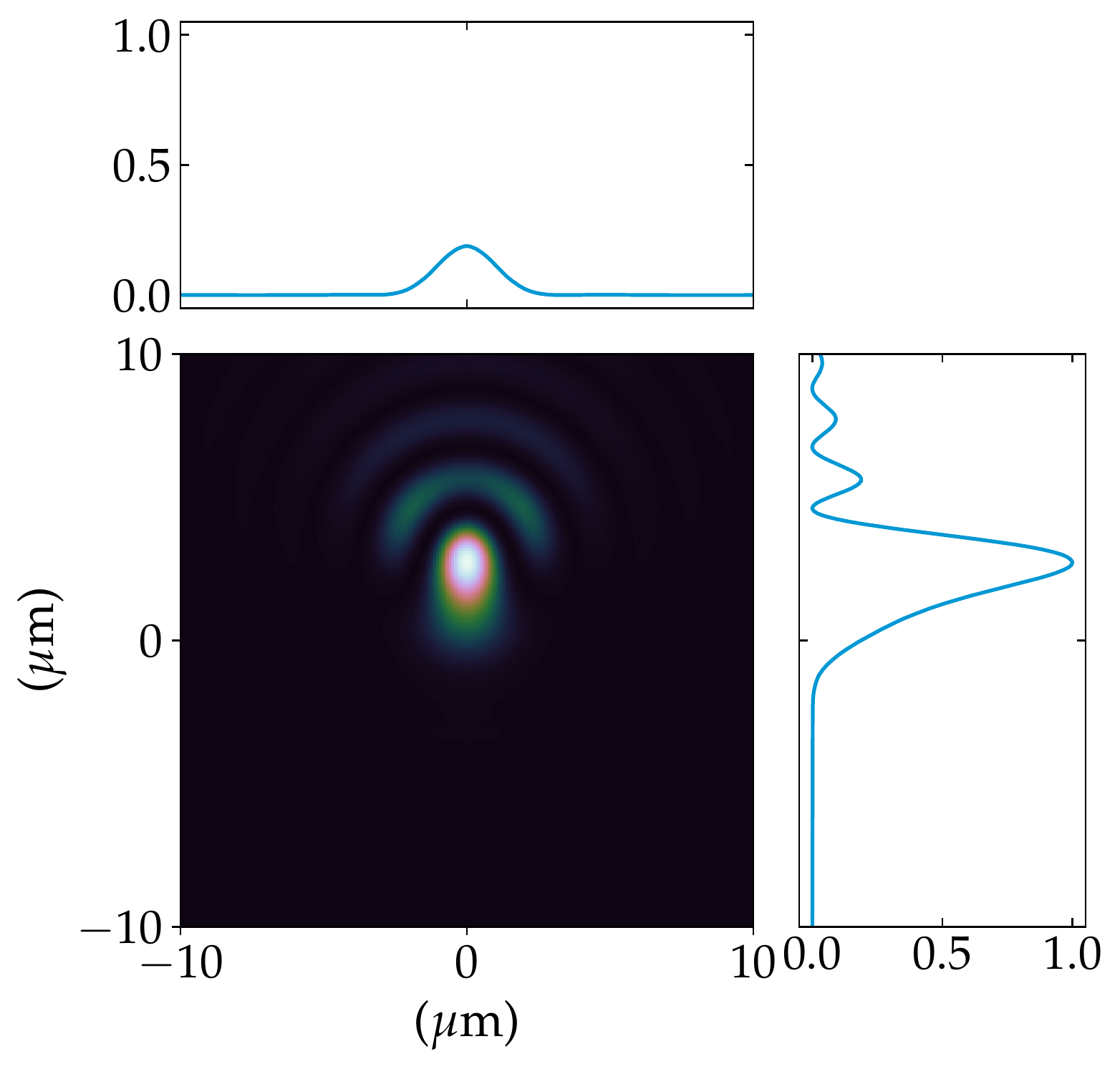}}\hspace{0.1cm}
        \subfloat[vertical caustics]{\includegraphics[height=3.5cm]{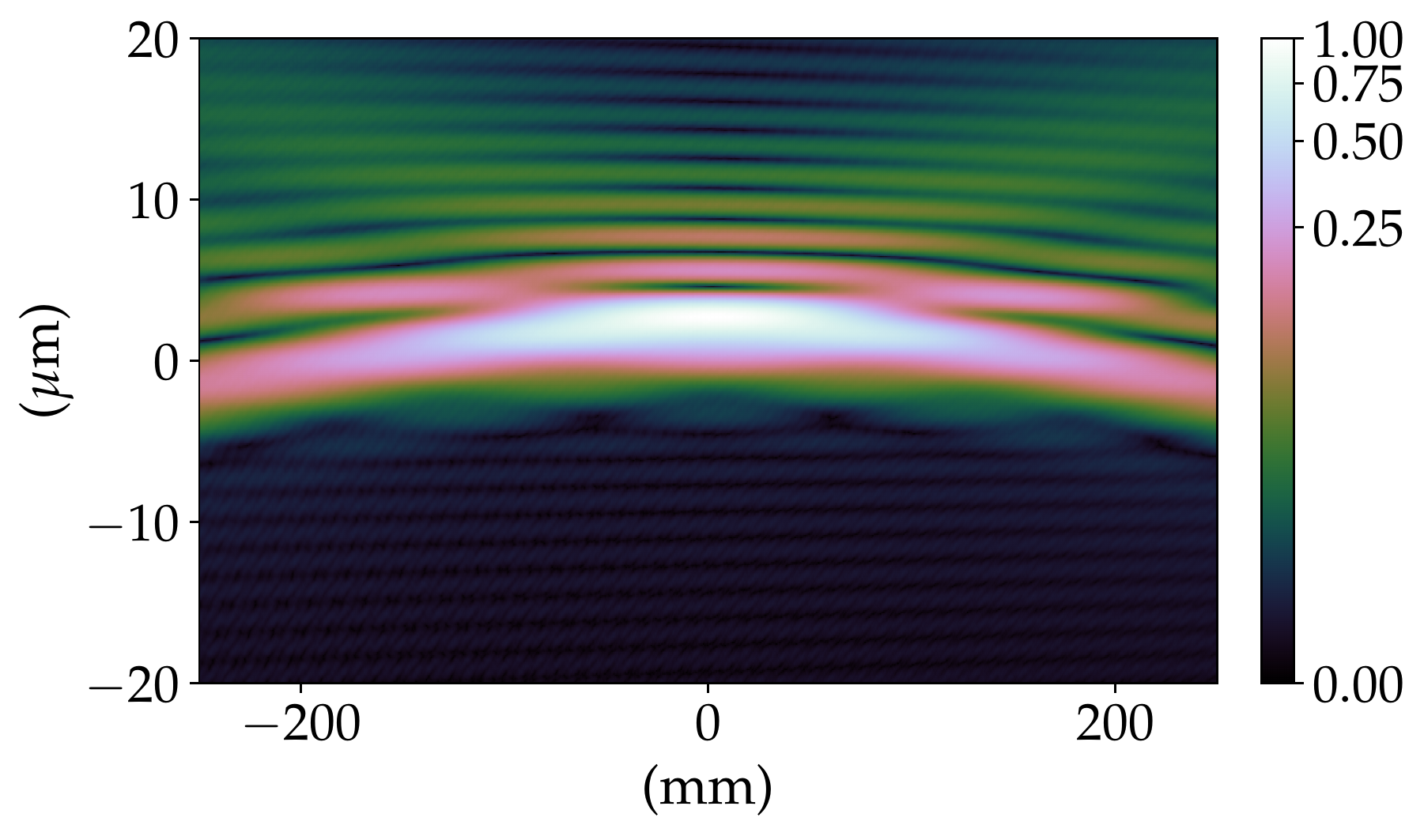}}
        \caption*{Anti-symmetric tilt of front- and back- surfaces by $\theta_{x_{\text{front}}}=-\theta_{x_{\text{back}}}=0.5^{\circ}$}
        \caption{\small Simulations of a lens with front and back focusing parabolic sections independently tilted and shown in Fig~\ref{fig:lens_cuts}(f) and (g). The (a) residual thickness, (b) point spread function  with cuts centred in $(0,0)$ and (c) the vertical beam caustics from -250~mm to 250~mm in respect to the focal plane of the symmetric tilt are shown on the top row. The (d) residual thickness, (e) point spread function and (f) the vertical beam caustics for the anty-symmetric case with the same conditions is presented in the bottom row. While the symmetric case has a residual phase proportional to the 4$^{\text{th}}$ power of the lateral coordinates in the direction of the tilt, elongating the beam focusing in the propagation direction and shifting it on the same direction - typical of spherical aberrations, the residual phase of the anti-symmetric case has residual phase proportional to the 3$^{\text{rd}}$ power of the lateral coordinates in the direction of the tilt and has a PSF typical of coma-aberrated systems.  2D-Beryllium lens with nominal radius $R=50~\mu\text{m}$, geometric aperture $A_{\diameter}=440~\mu\text{m}$ and $t_\text{wall}=20~\mu$m at 8~keV.} \label{fig:tilt_fs_CRL}
\end{figure}

When the axes of the parabolic front or/and back surfaces are not parallel to the optical axis, the lens active area appears to be tilted as shown in Figs.~\ref{fig:lens_cuts}(f)~and~(g). Similarly to what was introduced in the section \textit{Tilted lens} from \S\ref{sec:misalignments}~-~\textit{\nameref{sec:misalignments}}, both front and back surfaces are rotated according to the rotation matrices described in Eqs.~\ref{eq:affine} and the procedure described by Eq.~\ref{eq:affine2}. There are two subtle differences: the rotation angles from front and back surfaces can be chosen independently and rotation is only applied to the lens geometric aperture, and not to the whole front and back surfaces. The independent rotations allow for different regimes: one where both imprints are tilted with the same angle as in shown in Fig.~\ref{fig:lens_cuts}(f), which yields a residual phase similar to the one discussed in \textit{Tilted lens} from \S\ref{sec:misalignments}~-~\textit{\nameref{sec:misalignments}} and  one where there is no match of the front and back surfaces tilt, which yields an asymmetric residual phase as shown in Fig.~\ref{fig:tilt_fs_CRL}. By not applying the rotation to the region outside the lens geometric aperture, one avoids changing the lens projected thickness.

\subsection{Other sources of deviations from the parabolic shape}

So far, the modelling described here relies on translations and rotations of an ideal parabolic surface and investigating the residual phase. Another equally valid approach is to manipulate directly the residual phase and add it to the phase of an ideal focusing lens, which can be done fitting arbitrary surfaces or by introducing data from metrology of the optical element to be simulated.

\begin{figure}[t]
        \centering
        \subfloat[Be lens]{\includegraphics[height=2.7cm]{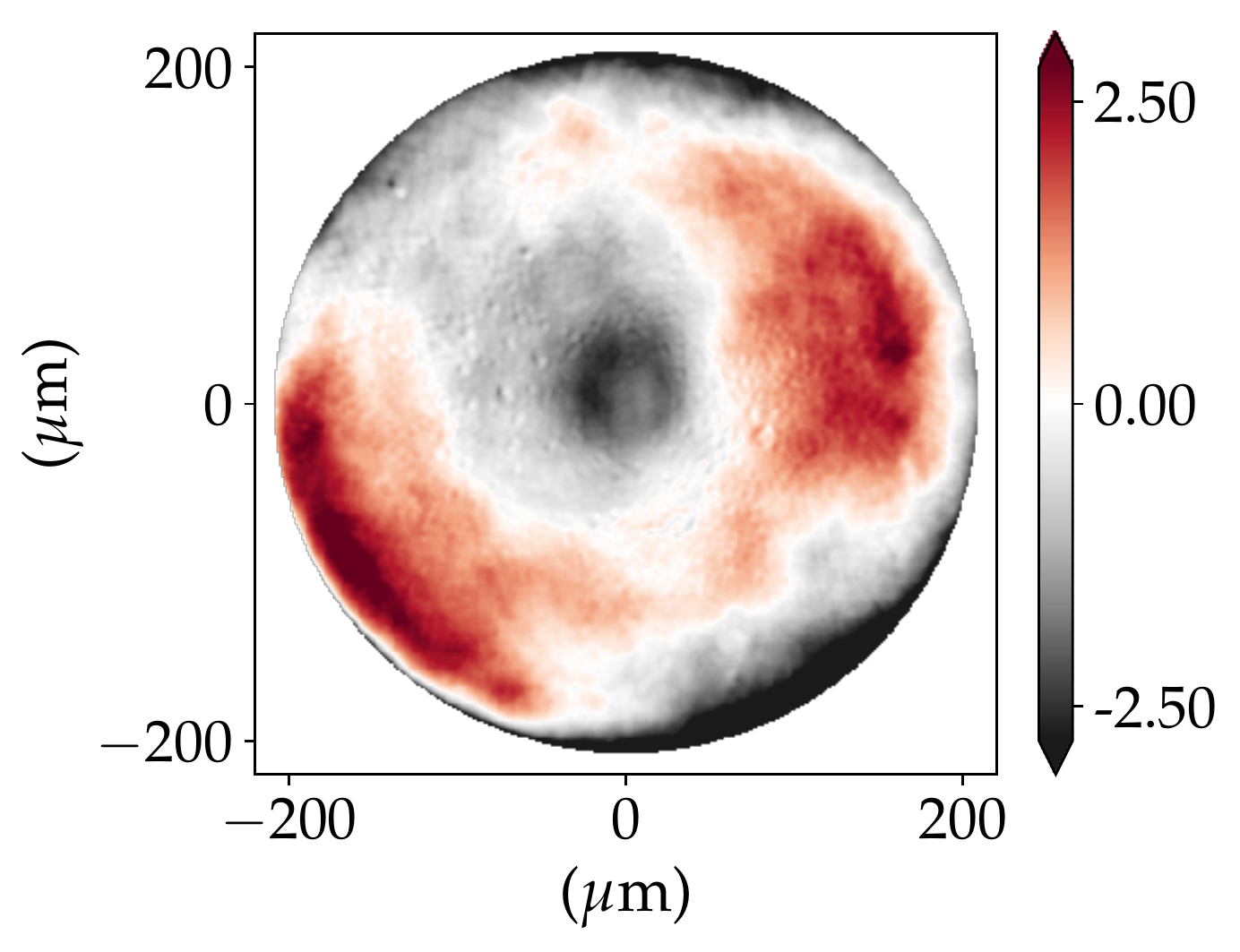}}\hspace{0.1cm}
        \subfloat[Polynomial decomposition]{\includegraphics[height=2.6cm]{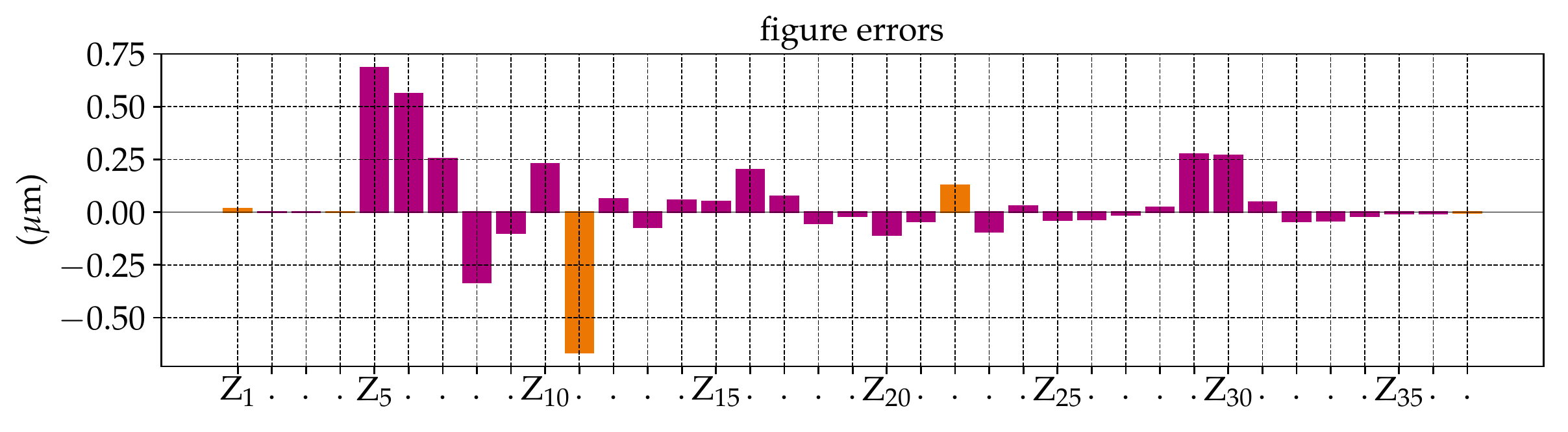}}\hspace{0.1cm}
        \subfloat[Reconstructed profile]{\includegraphics[height=2.7cm]{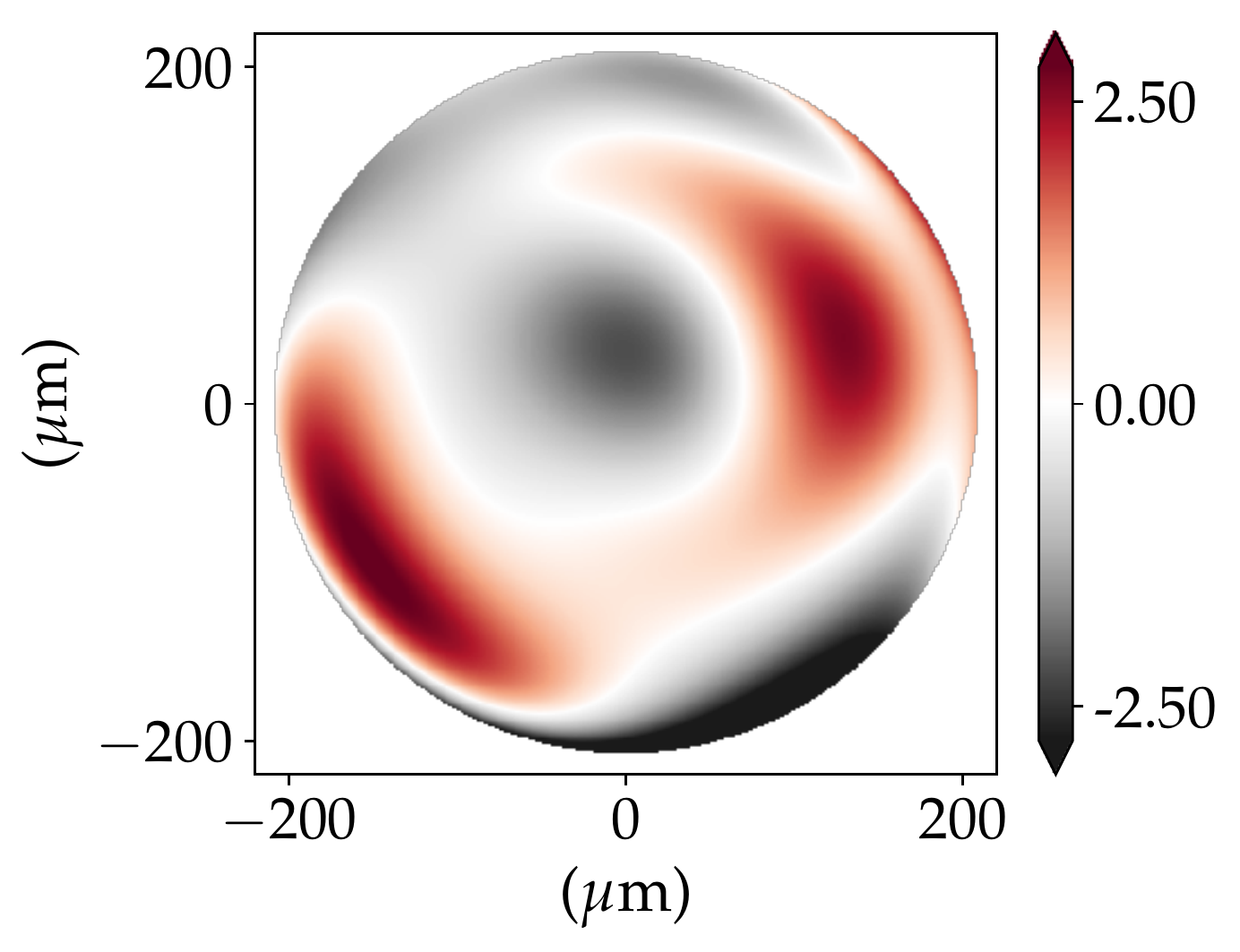}}\\
        \subfloat[Al lens]{\includegraphics[height=2.7cm]{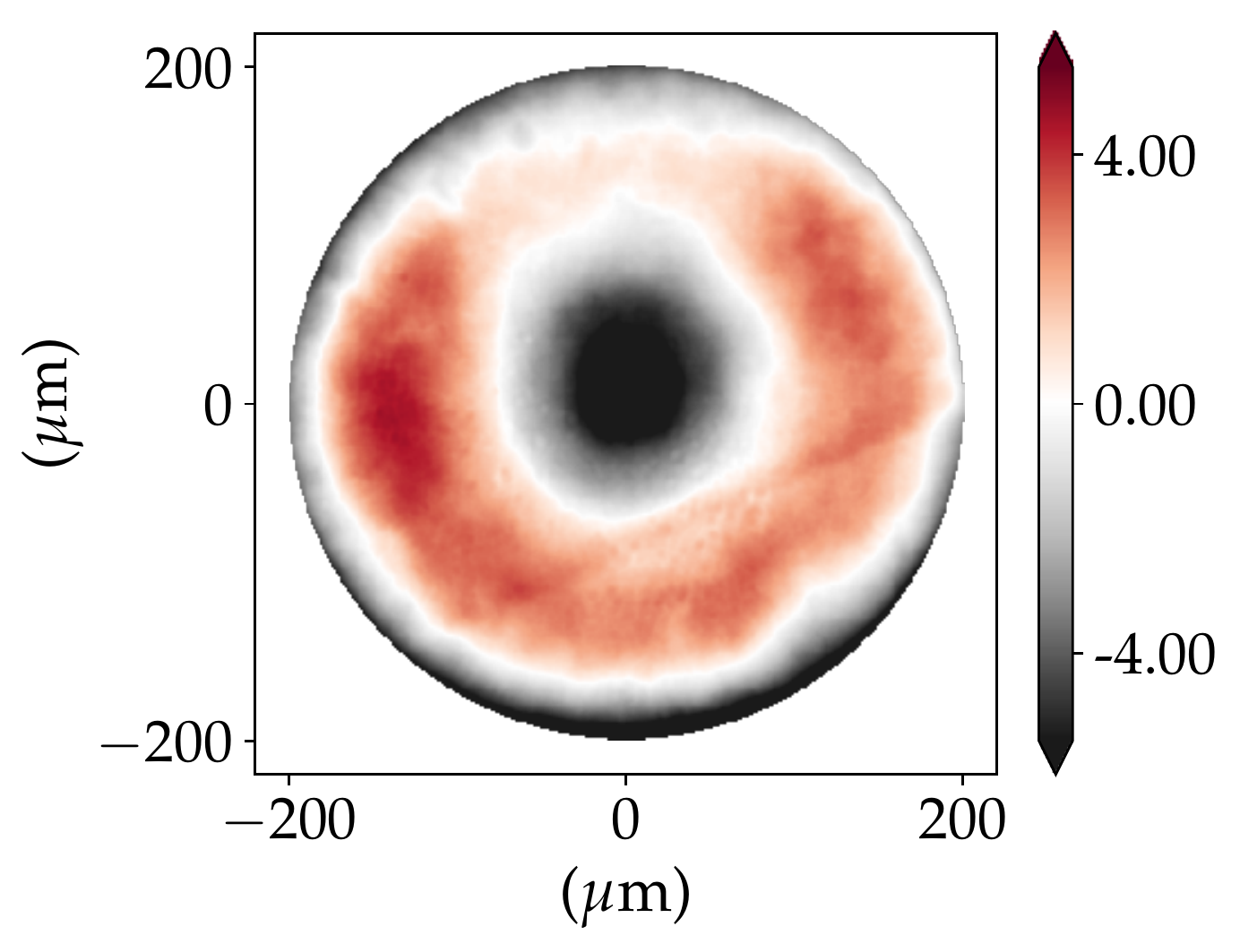}}\hspace{0.1cm}
        \subfloat[Polynomial decomposition]{\includegraphics[height=2.6cm]{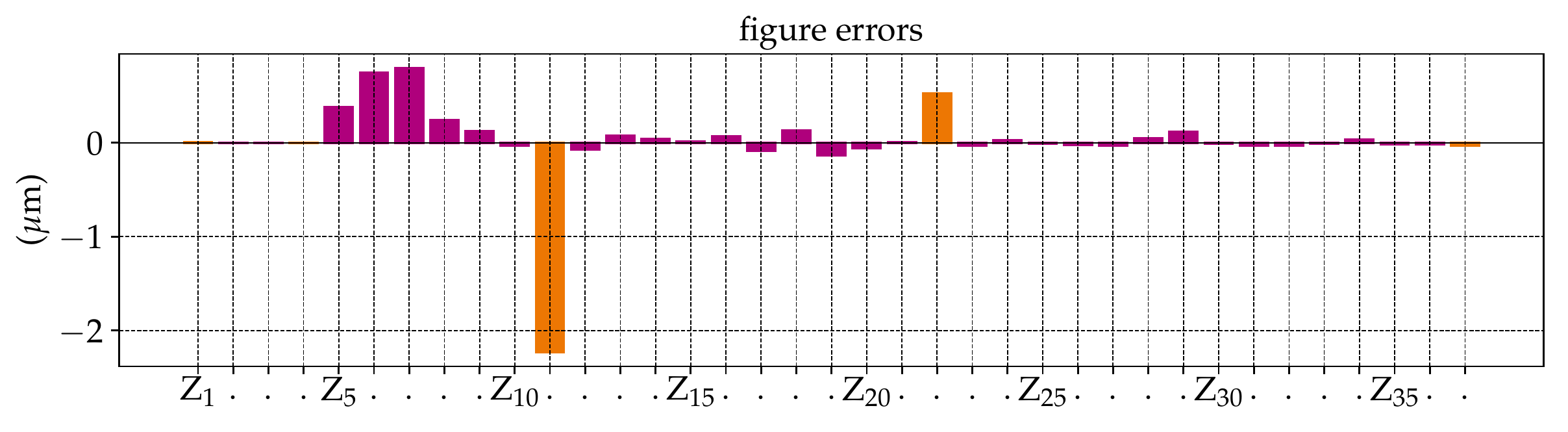}}\hspace{0.1cm}
        \subfloat[Reconstructed profile]{\includegraphics[height=2.7cm]{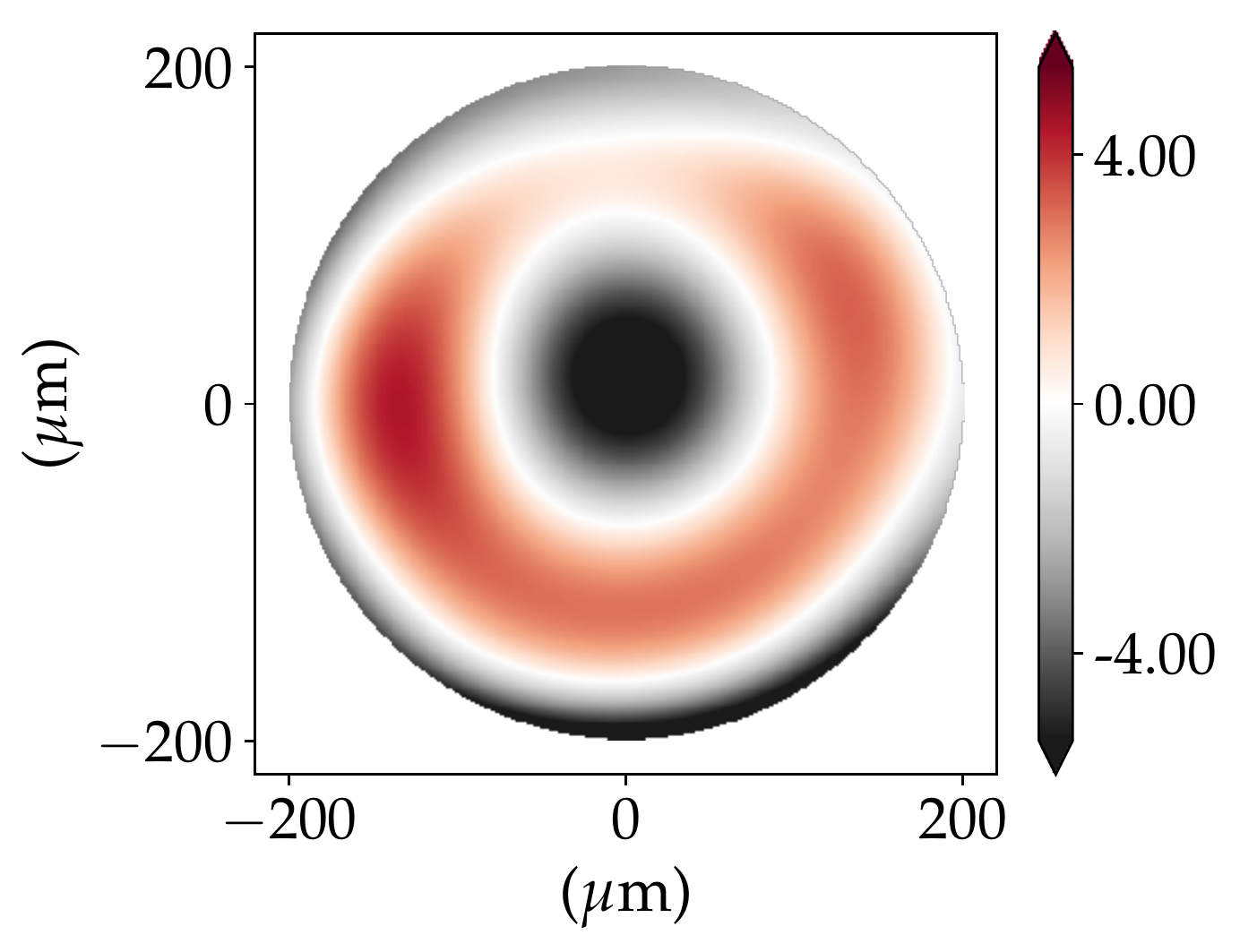}}\\
        \subfloat[Diamond lens]{\includegraphics[height=2.7cm]{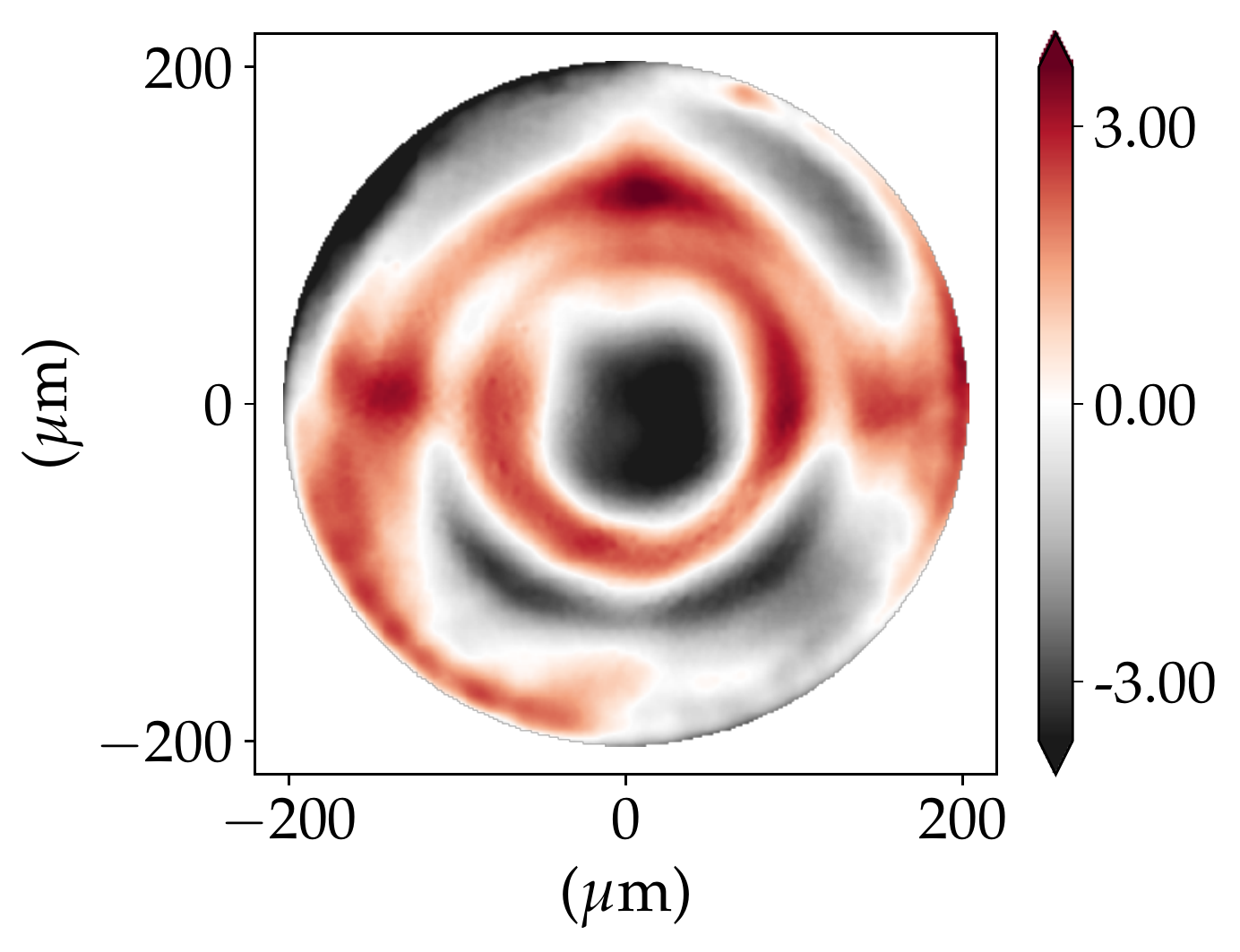}}\hspace{0.1cm}
        \subfloat[Polynomial decomposition]{\includegraphics[height=2.6cm]{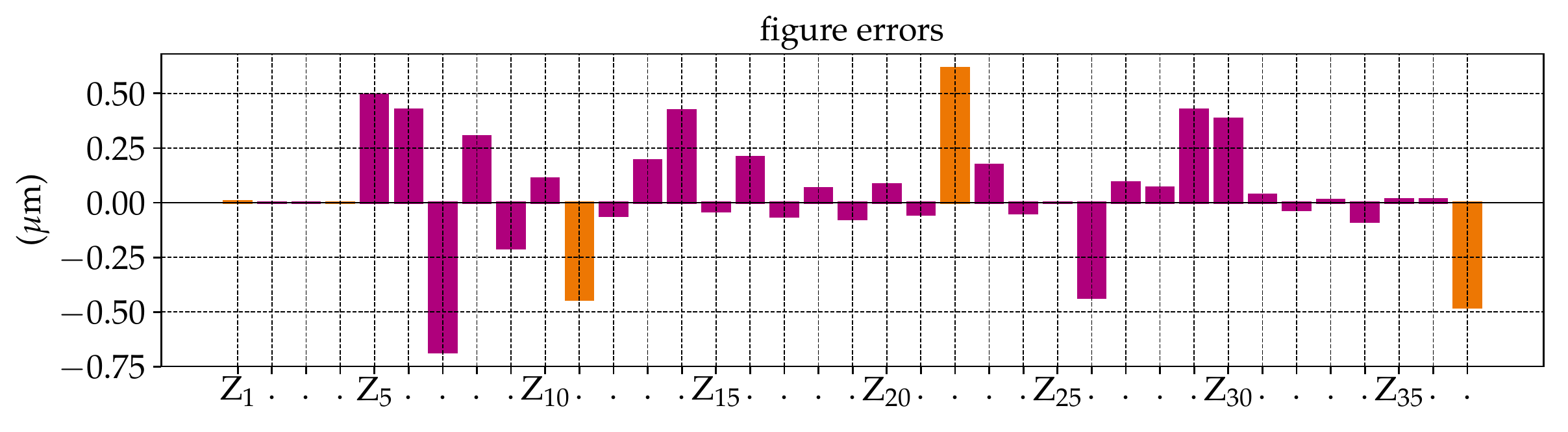}}\hspace{0.1cm}
        \subfloat[Reconstructed profile]{\includegraphics[height=2.7cm]{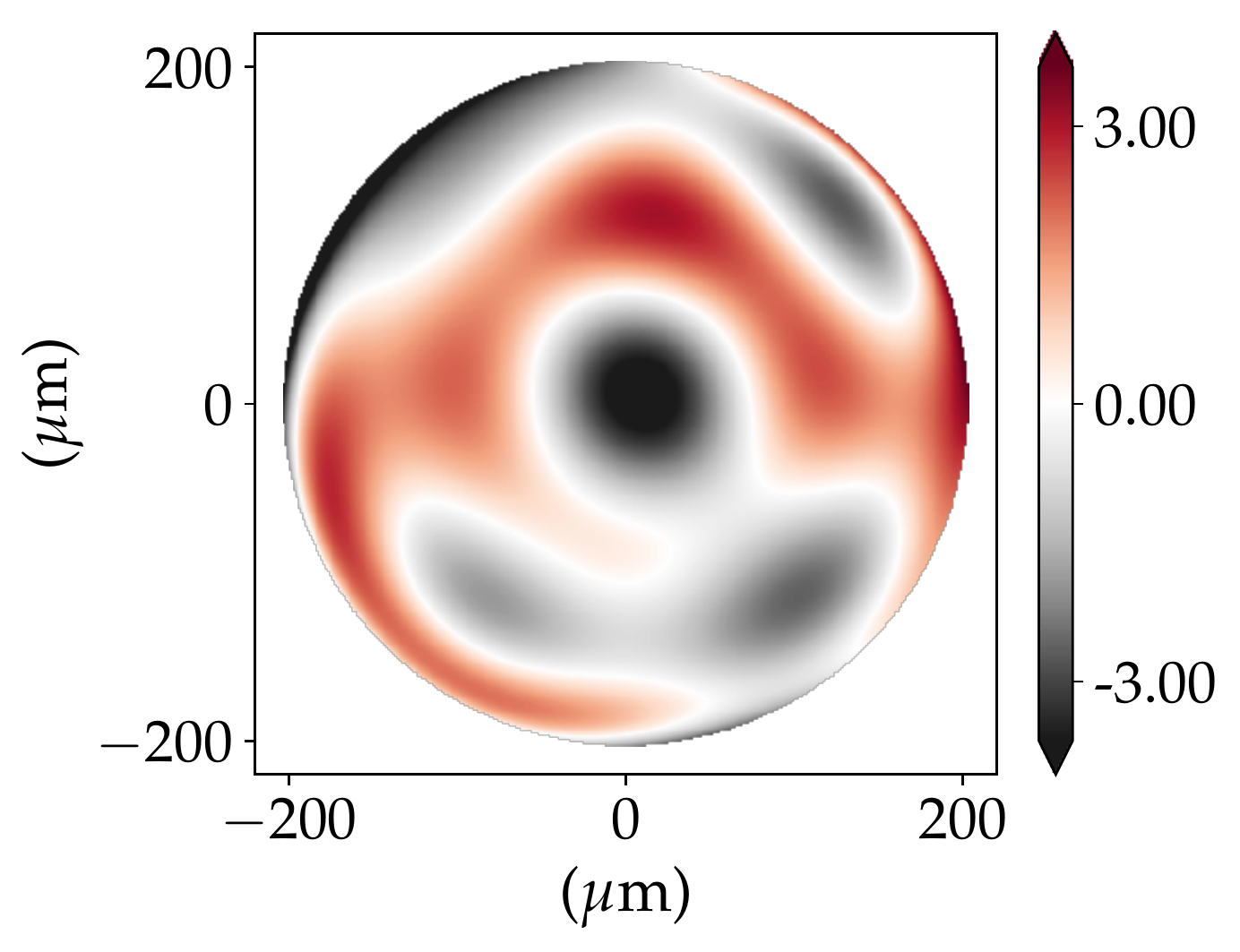}}
        \caption{\small \textbf{First row}: (a) accumulated profile with RMS value: $\sigma_z=1.4~\mu$m, (b) Zernike circle polynomial decomposition of the profile in (a) and (c) the reconstruction based on the coefficients of a 2D-Beryllium lens with nominal radius $R\sim47.3~\mu\text{m}$ and useful aperture $A_{\diameter}\sim417~\mu\text{m}$ with RMS value: $\sigma_z=1.3~\mu$m. \textbf{Middle row}: (d) accumulated profile with RMS value: $\sigma_z=2.6~\mu$m, (e) Zernike circle polynomial decomposition of the profile in (d) and (f) the reconstruction based on the coefficients of a 2D-Aluminium lens with nominal radius $R\sim46.2~\mu\text{m}$ and useful aperture $A_{\diameter}\sim396~\mu\text{m}$ with RMS value: $\sigma_z=2.5~\mu$m. \textbf{Bottom row}: (g) accumulated profile with RMS value: $\sigma_z=1.8~\mu$m, (f) Zernike circle polynomial decomposition of the profile in (g) and (h) the reconstruction based on the coefficients of a 2D-diamond lens with nominal radius $R\sim103.4~\mu\text{m}$ and useful aperture $A_{\diameter}\sim402~\mu\text{m}$ with RMS value: $\sigma_z=1.6~\mu$m.}
        \label{fig:metrology_zernike_profiles}
\end{figure}

\subsubsection*{Orthonormal polynomials}

A widespread form of representing optical aberrations of arbitrary shapes is by decomposing them into an orthonormal base. Perhaps the most ubiquitous set of aberration functions is given by the Zernike polynomials for a circular aperture, first described in [\cite{Zernike1934}]. Their appeal comes from the fact that not only they are directly related to Seidel (primary), Schwarzschild (secondary) and tertiary-aberrations\footnote{This jargon comes from a power-series expansion of the aberration function. There are five primary aberrations, nine secondary aberrations and fourteen aberration terms for the tertiary aberrations. They all involve spherical aberration, coma, astigmatism, field
curvature, distortion and variations of thereof [\cite{Mahajan2013}].} but also include piston and tilts; they form an orthonormal base, which means that the value of the coefficients is not affected by the removal of a particular term [\cite{Mahajan2007}]. Another advantage of the Zernike polynomial decomposition is that each orthonormal aberration coefficient is the standard deviation for that particular aberration over the exit pupil, which is valuable when evaluating the optical system compliance with the  Mar\'echal criteria and calculating the Strehl ratio [\cite{Mahajan1983}].

\begin{figure}[t]
        \centering
        \subfloat[PSF]{\includegraphics[height=5cm]{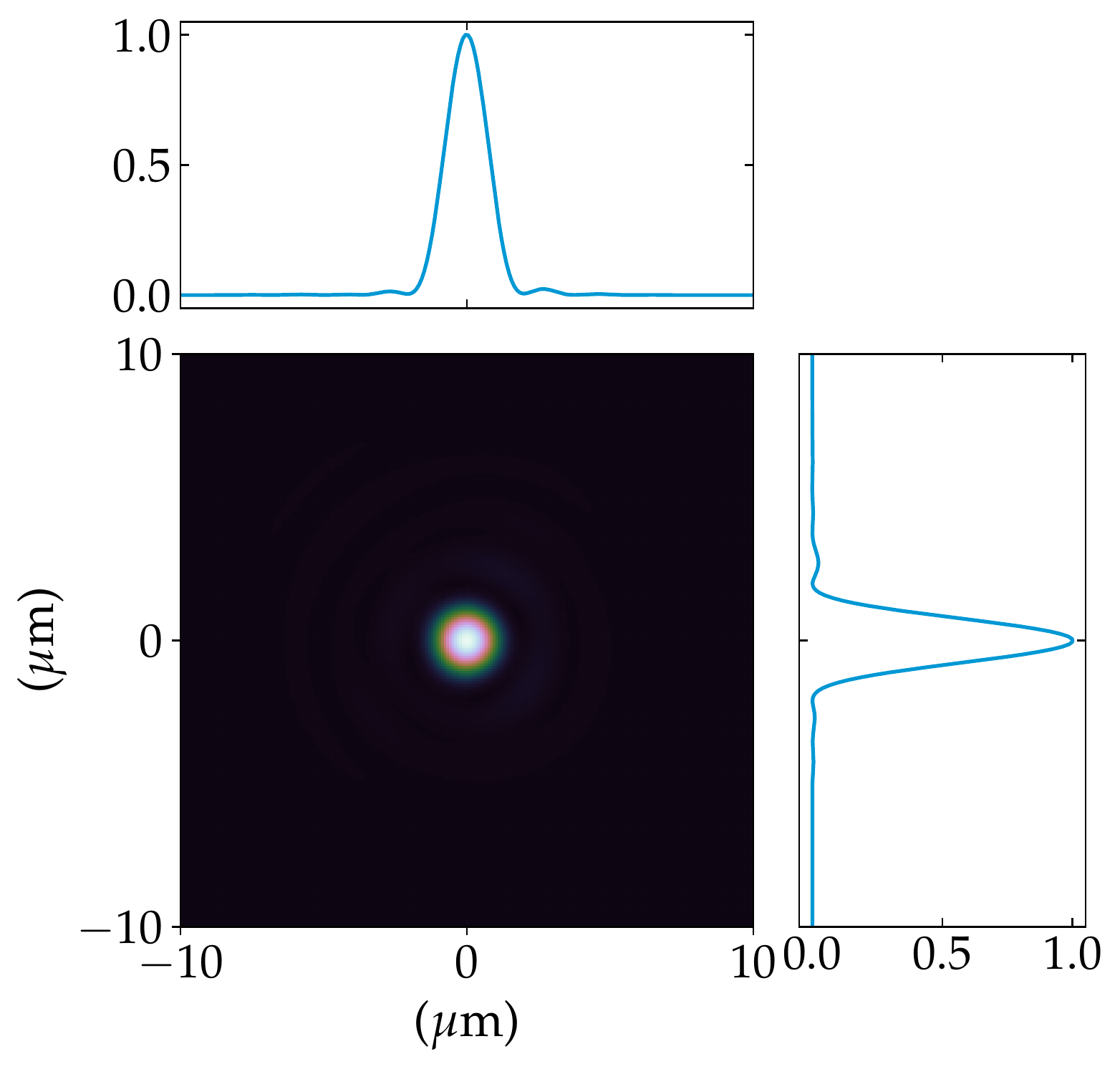}}\hspace{0.1cm}
        \subfloat[vertical caustics]{\includegraphics[height=3.5cm]{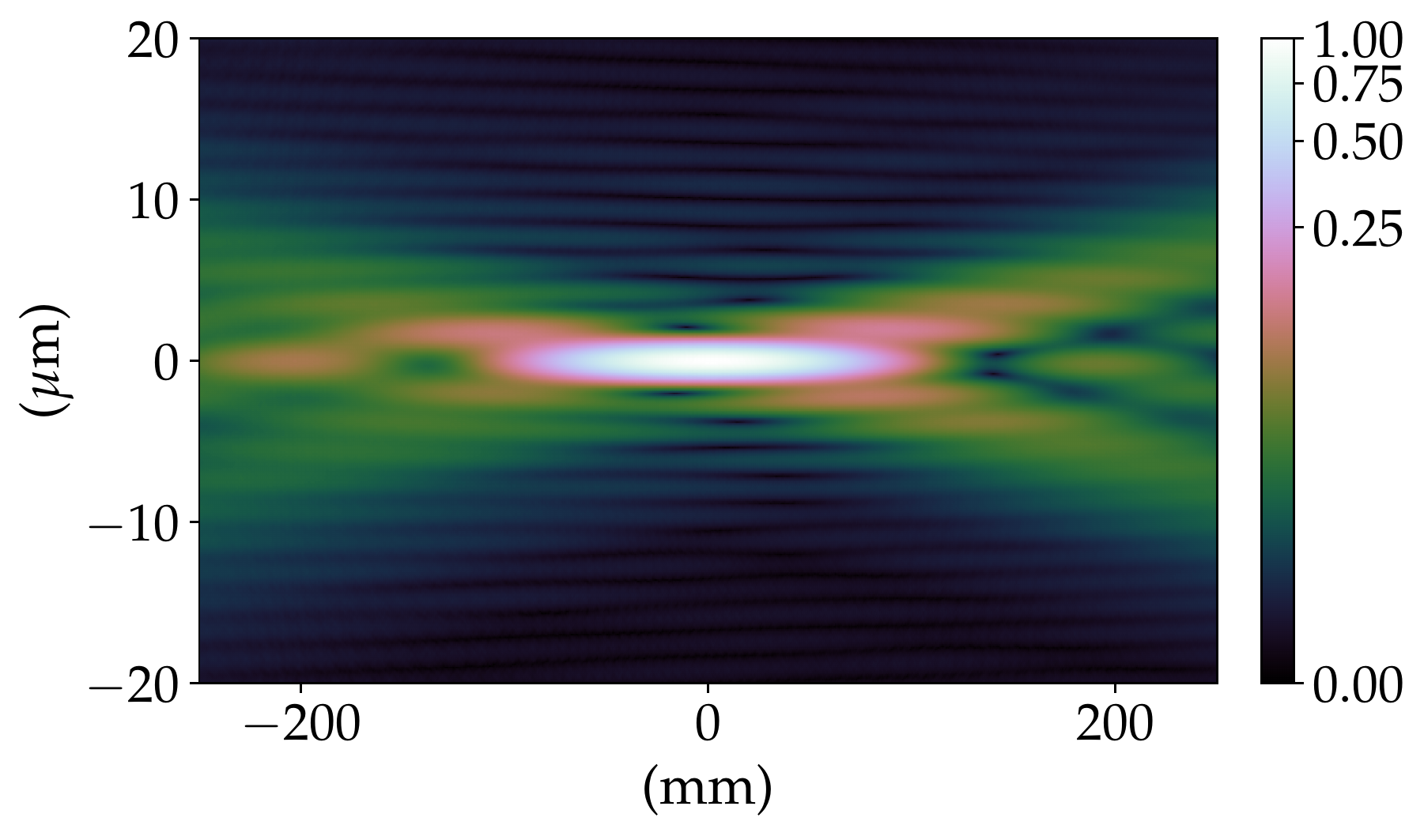}}\\\caption*{Simulations using the Zernike circle polynomials shown in Fig.~\ref{fig:metrology_zernike_profiles}(c)}
        \subfloat[PSF]{\includegraphics[height=5cm]{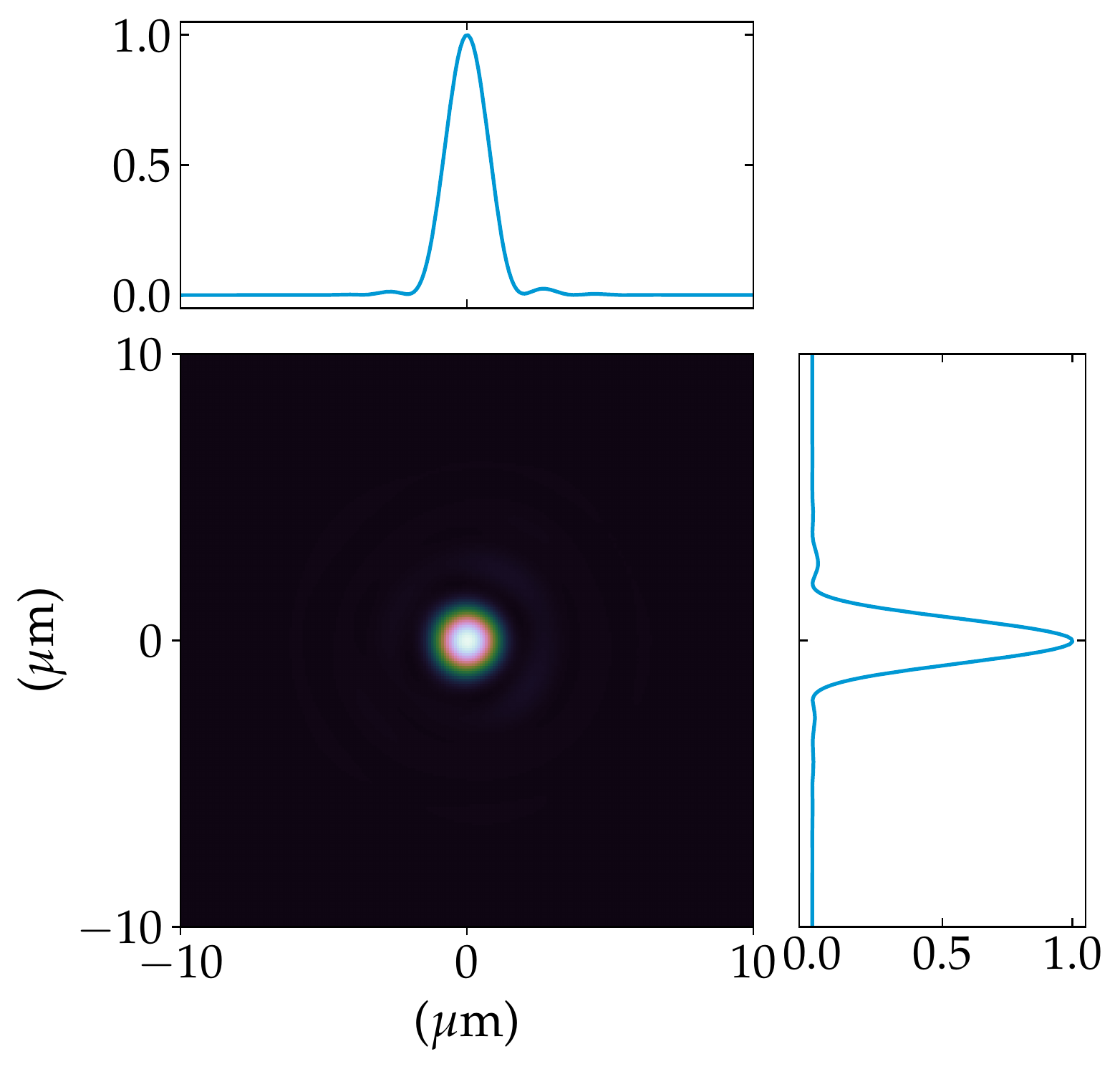}}\hspace{0.1cm}
        \subfloat[vertical caustics]{\includegraphics[height=3.5cm]{figures/cst_CRL_Be_metrology_intensity_cstc_Y_cstc_2D.pdf}}\\\caption*{Simulations using the metrology data shown in Fig.~\ref{fig:metrology_zernike_profiles}(a)}
        \caption{\small Simulations of a single 2D-Beryllium lens with nominal radius $R=50~\mu\text{m}$, geometric aperture $A_{\diameter}=440~\mu\text{m}$ and $t_\text{wall}=20~\mu$m at 8~keV. (a) and (c) point-spread function with cuts centred in $(0,0)$. (b) and (d) vertical beam caustics from -250~mm to 250~mm in respect to the focal plane. The difference between both profiles is almost not perceivable. This can be explained by the fact that the difference between the figure errors RMS value is almost negligible:  $\sigma_z=1.3~\mu$m (Zernike polynomial reconstruction) against $\sigma_z=1.4~\mu$m (metrology data). The impact in the the reduction in intensity at the focal position is almost negligible - cf. Mar\'echal criteria and Strehl ratio application to X-ray lenses and Eqs.~8 to 10 in [\cite{Celestre2020}] and Fig.~\ref{fig:Strehl}(c).}\label{fig:metrology_zernike_simualtions}
\end{figure}

For the aforementioned decomposition of the aberration function in an orthonormal base to retain its properties, the application of the circular Zernike polynomials must be limited to circular apertures. Other shapes of apertures with- or without obscuration can be obtained by Gram-Schmidt orthonormalisation and weighting of the Zernike circle polynomials [\cite{Swantner1994,Mahajan1995}]. X-ray optics systems often have a rectangular aperture and other two sets of polynomials are of particular interest in optical design: the set of orthonormal Zernike polynomials for a rectangular aperture  [\cite{Mahajan2007}] and the 2D-Legendre polynomial set for a rectangular aperture [\cite{Mahajan2010}]. Preferentially, the Zernike circle polynomials are applied to 2D focusing lenses with a circular aperture. For 2D focusing X-ray lenses with square aperture, low aspect ratio between horizontal and vertical apertures and not strongly astigmatic focusing, e.g. crossed planar X-ray lenses, the Zernike rectangular polynomials are preferred. The 1D focusing lens is better fit by the 2D Legendre polynomial set\footnote{Please, refer to [\cite{Ye2014}] for a comparison between 2D orthonormal sets for square apertures.}. Analysing and describing refractive X-ray optics using circular Zernike and 2D Legendre polynomials were first presented by [\cite{Koch2016}]. Profiles generated by Zernike circle polynomials are shown in the right-hand side of Fig.~\ref{fig:metrology_zernike_profiles} and their effect on a coherent X-ray beam in Fig.~\ref{fig:metrology_zernike_simualtions}(a) and (b).

\subsubsection*{Metrology data}

Any (unintentional) deviation of a parabolic shape can be considered as a source of manufacturing error. Each manufacturing process has some type of (signature) error associated to it and with the increasing number of exotic - or non-conventional - designs and tailored manufacturing strategies, it is beyond reasonable to create a model that could parametrise all sources of deviations from the parabolic shape. To circumvent that and to accurately model phase imperfection in compound refractive lenses, metrology data can also be used for optically imperfect X-ray lenses [\cite{Celestre2020, Chubar2020}]. Fig.~\ref{fig:metrology_zernike_profiles} shows three examples of lens figure errors from (a)-(c) a commercial pressed Beryllium lens, (d)-(f) an in-house pressed Aluminium lens and a (g)-(h) in-development laser-ablated Diamond lens from a commercial partner. The figure errors were measured with at-wavelength metrology and can be directly plugged into simulations as they describe the accumulated errors in projection approximation of both front and back focusing surfaces [\cite{Celestre2020, Berujon2020, Berujon2020a}]. The effect on a coherent X-ray of optical imperfections from metrology data is shown in Fig.~\ref{fig:metrology_zernike_simualtions}(c) and (d).

\begin{figure}[t]
        \centering
        \subfloat[misalignments]{\includegraphics[height=3.3cm]{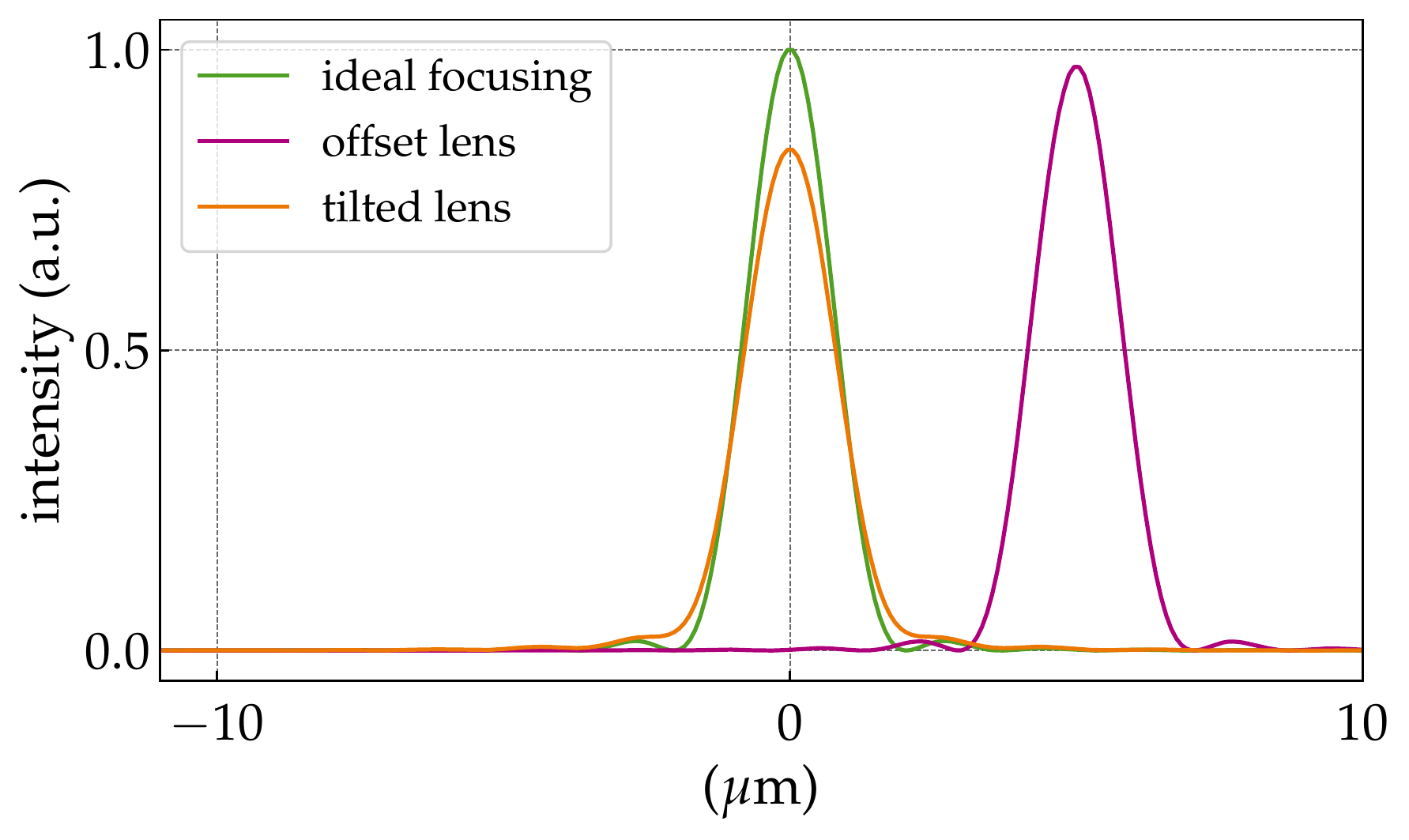}}\hspace{0.1cm}
        \subfloat[fabrication errors]{\includegraphics[height=3.3cm]{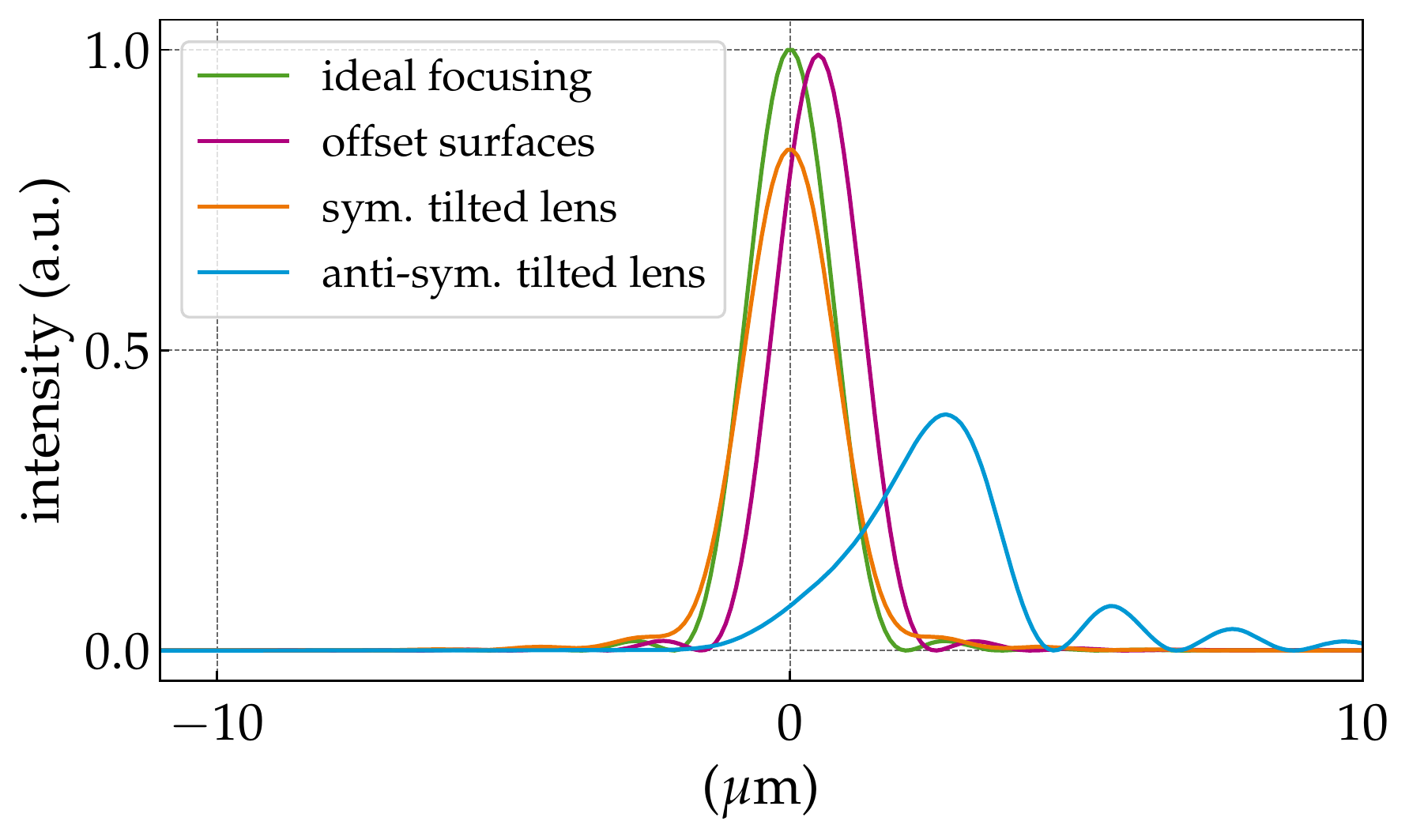}}\hspace{0.1cm}
        \subfloat[other sources or errors]{\includegraphics[height=3.3cm]{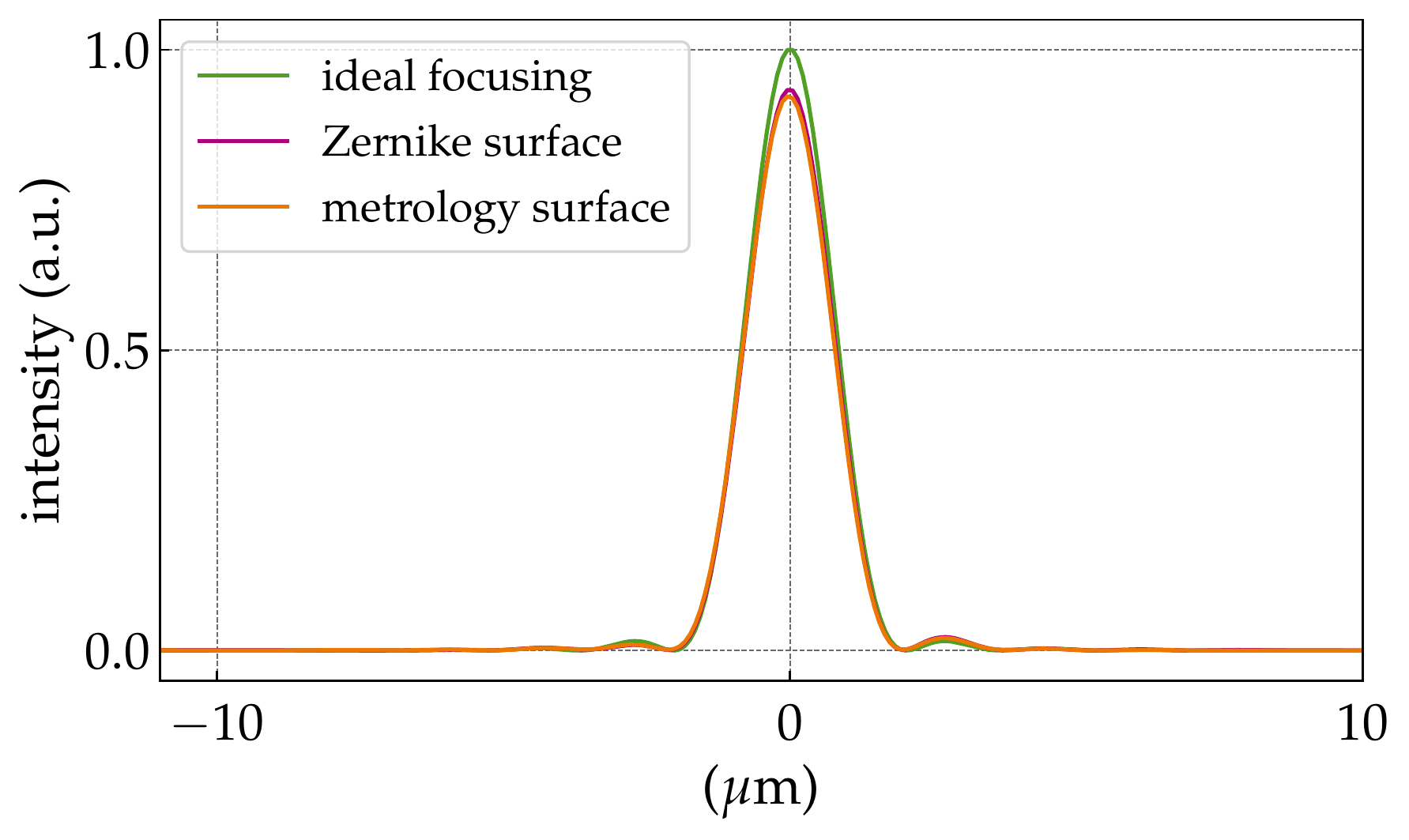}}
        \caption{\small Strehl ratio of the vertical cut at $x=0$ summarising the results from the diverse models presented (cf. Fig.~\ref{fig:shifted_CRL} to Fig.~\ref{fig:metrology_zernike_simualtions}).} \label{fig:Strehl}
\end{figure}

\section{IMPLEMENTATION}

The implementation of the modelling of the CRL, its misalignments and its figure errors is done using Python 3.7 and is fully compatible with the optical element class \texttt{SRWLOpt} described in the module \texttt{srwlib.py} from SRW\footnote{\url{github.com/ochubar/SRW}}. Each function representing either an X-ray lens or its figure errors returns a class \texttt{SRWLOptT} representing a generic transmission element storing amplitude transmission and optical path difference as a function of transverse coordinates. 

The main calculations for generating the X-ray lens transmission element is performed by the function \texttt{srwl\_opt\_setup\_CRL}. Apart from generating an ideal CRL (cf. \S\ref{sec:lens_ideal}~-~\textit{\nameref{sec:lens_ideal}}), this function implements the degrees of freedom discussed in \S\ref{sec:misalignments}~-~\textit{\nameref{sec:misalignments}}; longitudinal and transverse offsets as well as the tilts of the individual parabolic sections as described in \S\ref{sec:fabrication}~-~\textit{\nameref{sec:fabrication}}. The generation of the residual phase errors based on the polynomial expansion of the aberration function in the exit pupil is done by \texttt{srwl\_opt\_setup\_CRL\_errors}. To calculate the aberration function, the user can either use a list with the polynomial coefficients or enter an RMS value for their sum (equivalent to fixing the piston value), in which case, they will be randomly calculated will comply with the RMS value limited by the user input. This function generates the residual errors and should be used in conjunction with \texttt{srwl\_opt\_setup\_CRL} to simulate an aberrated focusing lens. The aberration functions in \texttt{srwl\_opt\_setup\_CRL\_errors}\footnote{The functions used by \texttt{srwl\_opt\_setup\_CRL\_errors} to generate the 2D circular Zernike polynomials contain pieces of codes from the module \texttt{libtim-py} from Tim van Werkhoven, that had to be brought to Python 3.7 and in some places, small bugs fixed - this module has a Creative Commons Attribution-Share Alike
license. The 2D rectangular Zernike was originally inspired by the analytical formulation from the module \texttt{opticspy} from Xing Fan - this module has an MIT license. The formulation from the module was based on the equations from [\cite{Mahajan2007}], but had to be corrected with the errata published in [\cite{Mahajan2012}].} use the analytical solutions from [\cite{Mahajan2007}] with the corrections from [\cite{Mahajan2012}] and the solutions from [\cite{Mahajan2010}]. 

The generation of a surface based on the metrology data is done by \texttt{srwl\_opt\_setup\_CRL\_metrology}. The metrology data should be saved as an ASCII file (.dat) as defined by the function \texttt{srwl\_uti\_save\_intens\_ascii} from the module \texttt{srwlib.py}. The function \texttt{srwl\_opt\_setup\_CRL\_metrology} can be used to simulate figure errors, in which case, much like \texttt{srwl\_opt\_setup\_CRL\_errors} it requires the use of \texttt{srwl\_opt\_setup\_CRL} or it can be used to simulate a full measured profile. 

This library extension is currently available under a CC BY-SA 4.0 license at the \href{https://gitlab.esrf.fr/celestre/barc4RefractiveOptics}{barc4RefractiveOptics} GitLab repository\footnote{\url{gitlab.esrf.fr/celestre/barc4RefractiveOptics}}, where more information on the implemented functions can be found.

\section{CONCLUSION}

Based on the ideal CRL modelling present in SRW [\cite{Baltser2011}], we presented an expanded model that allows the CRL more degrees of freedom, which allows accounting for typical phase errors encountered in X-ray lenses. We did that by decoupling the front and back focusing and allowing them to be independently tilted or offset. A global tilt and offset are also applied to the lenses. For the case where the simple tilt and offset of the lens front and surfaces is not enough, we proposed the use of a set of orthonormal polynomials for 2D and 1D focusing lenses with circular and rectangular apertures. If available, metrology data can also be converted into a transmission element and be used in SRW simulations as described in [\cite{Celestre2020}]. The results from Fig.~\ref{fig:shifted_CRL} to Fig.~\ref{fig:metrology_zernike_simualtions} are summarised in Fig.~\ref{fig:Strehl}.

The codes are publicly available under a CC BY-SA 4.0 license at the \href{https://gitlab.esrf.fr/celestre/barc4RefractiveOptics}{barc4RefractiveOptics} repository. The calculations of the thickness in projection approximation (1D and 2D) are done outside the main function \texttt{srwl\_opt\_setup\_CRL}, the idea being that such modelling could be used in other wave-propagation codes if there is interest. The CRL modelling is available in barc4RefractiveOptics until the eventual merge with the official SRW repository. Future releases of barc4RefractiveOptics should include aspheric surfaces [\cite{SanchezdelRio2012}], Zernike-Gauss polynomials for apodised apertures [\cite{Mahajan1995}], a function for generating a random surface based on the power spectrum density (PSD) of a lens surface and be completely backwards compatible with the present CRL model from SRW, that is, include the polychromatic field calculation and inhomogeneities in the material as introduced in [\cite{Roth2014}]. 

With increased activities in beamline design and new opportunities for wave-front preserving optics, we expect that the modelling presented here can be used to increase accuracy in beamline design in new high-energy synchrotrons and X-FELs. Tolerancing and manufacturing of optical elements and wavefront correction plates design are also areas that should profit from more accurate modelling of CRL.

\acknowledgments 
 
R.C. and O.C. acknowledge the "DOE BES Field Work Proposal PS-017 funding". R.C. thanks Prof. V. Mahajan (University of Arizona) and Prof. H. Gross (University of Jena) for the discussions on orthonormal polynomials in wavefront analysis and pointing out relevant literature. R.C. and T.R. thank C. Detlefs from ID06-ESRF for allowing to use some of the ESRF-EBS beamlines commissioning time for measuring the lenses presented in Fig.~\ref{fig:metrology_zernike_profiles}.
\small
\bibliography{report} 
\bibliographystyle{spiebib} 

\end{document}